\documentclass[prd , aps,12 pt,nofootinbib,superscriptaddress, preprintnumbers,notitlepage]{revtex4-2}
\pdfoutput=1
\usepackage{graphicx}
\usepackage{amssymb}
\usepackage{mathtools}
\usepackage{epstopdf}
\usepackage{color}
\usepackage{bbold}
\usepackage{float}
\usepackage[title]{appendix}
\usepackage[colorlinks, citecolor=blue, linkcolor=black, urlcolor=purple]{hyperref}
\usepackage{amsfonts}
\usepackage{amsmath}
\usepackage{setspace}
\usepackage{epsfig}
\usepackage[caption=false]{subfig}
\usepackage[usenames, dvipsnames]{xcolor}
\usepackage{tabularx}
\usepackage{empheq}
\usepackage{mathrsfs}
\usepackage{xcolor}
\usepackage{ulem}

\usepackage{units}

\newcommand{\be}{\begin{eqnarray}}
\newcommand{\ee}{\end{eqnarray}}
\newcommand{\bea}{\begin{eqnarray}}
\newcommand{\eea}{\end{eqnarray}}
\newcommand{\nn}{\nonumber}

\newcommand{\beq}{\begin{equation}}
\newcommand{\eeq}{\end{equation}}

\definecolor{bluDT}{cmyk}{1,0.5,0,0.3}

\definecolor{darkblue}{rgb}{0.2,0.2,0.9}
\definecolor{colorRTD}{rgb}{.2,.2,.7}

\definecolor{colorHD}{rgb}{.2,0.9,.0.9}

\usepackage{soul}

\begin{document}

\preprint{CERN-TH-2021-138}

\title{\LARGE Sliding Naturalness:\\
Cosmological Selection of the Weak Scale
}

\author{Raffaele Tito D'Agnolo}
\affiliation{Université Paris-Saclay, CEA, Institut de physique théorique, 91191, Gif-sur-Yvette, France}
\author{Daniele Teresi}
\affiliation{CERN, Theoretical Physics Department, 1211 Geneva 23, Switzerland}


\begin{abstract}
We present a cosmological solution to the electroweak hierarchy problem.
After discussing general features of cosmological approaches to naturalness,  we extend the Standard Model with two light scalars very weakly coupled to the Higgs and present the mechanism, which we recently introduced in a companion paper to explain jointly the electroweak hierarchy and the strong-CP problem. In this work we show that this solution can be decoupled from the strong-CP problem and discuss its possible implementations and phenomenology. The mechanism works with any standard inflationary sector, it does not require weak-scale inflation or a large number of e-folds, and does not introduce ambiguities related to eternal inflation. 
The cutoff of the theory can be as large as the Planck scale, both for the Cosmological Constant and for the Higgs sector. Reproducing the observed dark matter relic density fixes the couplings of the two new scalars to the Standard Model, offering a target to future axion or fifth force searches. Depending on the specific interaction of the scalars with the Standard Model, the mechanism either yields rich phenomenology at colliders or provides a novel joint solution to the strong-CP problem. We highlight what predictions are common to most realizations of cosmological selection of the weak scale and will allow to test this general framework in the near future.

\end{abstract}

\maketitle
\onecolumngrid

\tableofcontents
\newpage

\section{Introduction} 
The questions surrounding the Higgs boson mass have driven most of the research in particle physics in the last decades. Experiments at LEP and at the LHC have neither discovered the symmetries that we expected~\cite{Dimopoulos:1981zb, Dimopoulos:1981yj, Bagger:1990qh, Martin:1997ns, Weinberg:2000cr, tHooft:1979rat, Dimopoulos:1979es, Terazawa:1976xx, Kaplan:1983fs, Kaplan:1983sm, Dugan:1984hq} nor those that initially we did not expect~\cite{Chacko:2005pe, Burdman:2006tz}, leaving the value of the Higgs mass as puzzling as ever.

This situation has led some to question the problem rather than its proposed solutions. However, the problem is more concrete and interesting today than it ever was. It is more concrete because we have discovered the Higgs boson, measured its mass and established that it is a fundamental scalar\footnote{At least up to a factor of ten in energy above its mass}. The results from LEP were already pointing to a naturalness problem, but before the LHC we did not know what caused electroweak symmetry breaking in the Standard Model. 

The problem is now more interesting because its most elegant solutions can not be realized in their simplest form and it is unclear whether we should abandon them entirely and radically change our outlook on the weak scale or accept some amount of tuning as a fundamental aspect of physics. Either way we will learn something new about Nature. 

Possibly the most fascinating aspect of this question is that even ignoring it amounts to making important assumptions about physics at high energies. 
The Higgs boson mass is not calculable in the Standard Model, it is a measured parameter of the effective theory, so we could say that in our current description of Nature there is no problem and forget the whole issue. However this leaves open only two possibilities: 1) The Higgs mass is not calculable at any energy 2)  There is no mass scale beyond the Standard Model sufficiently strongly coupled to the Higgs to generate a fine-tuning problem. The first option, even if seemingly harmless, strongly constrains fundamental physics at high energies, to the point that we do not know a theory of quantum gravity that realizes it. The second one has interesting implications for model building and the description of other aspects of fundamental physics (dark matter, gauge coupling unification, ...)~\cite{Farina:2013mla,deGouvea:2014xba,Hambye:2018qjv}, and it forces us to think about theories of gravity with no new scales~\cite{Stelle:1976gc,Salvio:2014soa, Giudice:2014tma, Kannike:2015apa, Salvio:2017qkx} whose consistency is still unclear~\cite{Lee:1969fy,Salvio:2015gsi,Strumia:2017dvt,Gross:2020tph}. 

At the moment the (theoretically) most conservative attitude is to assume that supersymmetry (or anything else that makes the Higgs mass calculable) exists below the scale of quantum gravity. For concreteness we can imagine that string theory describes gravity at high energies and supersymmetry is broken somewhere below the string scale. In this case, at the theory level the naturalness problem of the Higgs mass squared can be stated sharply, already at tree-level and without any ambiguity. The Higgs mass is a calculable function of supersymmetric parameters that in principle we can measure independently. If two or more measured contributions to the Higgs mass are much larger in absolute value than its central value we want to understand why.  It is not guaranteed that the explanation will manifest itself at low energy, it might be related to the distribution of supersymmetry breaking parameters in a Multiverse or to the constraints imposed on their values by quantum gravity. However, even in these cases, thinking about the problem can shed light on fundamental aspects of physics. 

We have been looking for symmetric (or dynamical) explanations for the Higgs mass for more than 40 years and we have not yet found any obvious sign that they are realized in Nature. This has generated a ``little" hierarchy problem~\cite{Barbieri:1987fn, Barbieri:2000gf}. We have established a hierarchy between the Higgs mass $m_h$ and the scale at which new sources of flavor and CP violation can appear in Nature. This considerably complicates extending the SM to accommodate a symmetry or new dynamics that can protect the Higgs mass. The problem is further complicated by the null direct searches at LEP and the LHC.

Faced with these results we can take a different perspective and consider seriously the existence of a landscape for $m_h^2$. If we accept the existence of a vast landscape of vacua (for instance because of the cosmological constant or just because of string theory), it is likely that $m_h^2$ varies from vacuum to vacuum. Note that even if we extrapolate to the extreme the explanatory power of current swampland conjectures~\cite{Palti:2019pca} and imagine that the measured Cosmological Constant (CC) can be understood from the internal consistency of string theory, we still expect the existence of a vast landscape of vacua. 

Historically the existence of a landscape for $m_h^2$ coincides with anthropic solutions to the electroweak hierarchy problem~\cite{Agrawal:1997gf}.  Recently a new class of ideas emerged that makes a very different use of the landscape~\cite{Graham:2015cka, Arkani-Hamed:2016rle, Giudice:2019iwl, Strumia:2020bdy, Csaki:2020zqz, TitoDAgnolo:2021nhd}, with much better prospects for detection and little or no recourse to anthropic arguments. In these models a dynamical event is triggered by the Higgs Vacuum Expectation Value (vev) during the early history of the Universe. This event selects the value of $m_h^2$ that we observe today, leaving traces at low energy that can escape current searches, but are in principle detectable in the near future.

Here we discuss a proposal in this class with the following qualities: 1) It is entirely described by a simple polynomial potential for two weakly-coupled light scalars 2) it does not make any assumption on what can explain current CMB observations, in particular it is compatible with one's favorite mechanism (and scale) for inflation, but also with de Sitter swampland conjectures 3) it can explain a small value of the Higgs vev $v\simeq 246$~GeV, even if the Higgs is coupled at $\mathcal{O}(1)$ with particles at $M_{\rm Pl}$, 4) it is not affected by problems of measure in the landscape\footnote{If the landscape is populated via eternal inflation there will be a measure problem if one is interested in understanding what values of fundamental parameters are more likely in the Multiverse. However this does not affect the validity of the mechanism, since we are not asking probabilistic questions in the Multiverse. We instead have a theory where all unwanted patches are either always empty or always crunch. So we never need to know if the unwanted patches are more or less likely (occupy a smaller or larger volume in the Multiverse) than the one that we observe.}. 

In a companion paper~\cite{TitoDAgnolo:2021nhd} we have already discussed one realization of this idea that simultaneously explains the value of the Higgs boson mass and of the QCD $\theta$-angle. Here we discuss the general features of this mechanism, what are the possible implementations and their phenomenology. Furthermore, we describe how this idea compares to other ideas that trace the origin of the weak scale to the early history of the Universe. The idea of crunching away ``unwanted" patches of the Multiverse, that we exploit in this work, was already discussed in relation to fine-tuning problems in~\cite{Bloch:2019bvc,Strumia:2020bdy,Csaki:2020zqz}. 

In Section~\ref{sec:CN} we discuss general features of cosmological naturalness that place the predictions of our mechanism in a broader context and highlight the common predictions of these mechanisms, which can be used to experimentally test this framework. In Section~\ref{sec:mechanism} we describe the basic idea behind our proposal. In Section~\ref{sec:pheno} we discuss the cosmology of the mechanism and its predictions for dark matter. In Section~\ref{sec:H1H2trigger} and ~\ref{sec:GGtrigger} we describe two operators that couple the new scalars to the SM and their phenomenology: the first one yields a rich phenomenology at colliders, the second one allows to solve also the strong-CP problem in a novel way~\cite{TitoDAgnolo:2021nhd}. 
We conclude in Section~\ref{sec:conclusion}. 

\section{General Features of Cosmological Naturalness} \label{sec:CN}

 A number of creative ideas that trace the origin of the weak scale to early times in the history of the Universe are present in the literature~\cite{Dvali:2003br,Dvali:2004tma,Graham:2015cka,Arkani-Hamed:2016rle, Arvanitaki:2016xds,Geller:2018xvz,  Cheung:2018xnu, Giudice:2019iwl, Strumia:2020bdy, Csaki:2020zqz, Arkani-Hamed:2020yna,  Giudice:2021viw,TitoDAgnolo:2021nhd}. Taken at face value these ideas seem widely different, selecting the weak scale by unrelated mechanisms and predicting different phenomenology. In this Section we identify the basic structure common to these proposals and find that a large subset of these ideas have common ingredients which often lead to similar low-energy predictions.  
 
\begin{figure}[t]
\includegraphics[width=0.45\textwidth]{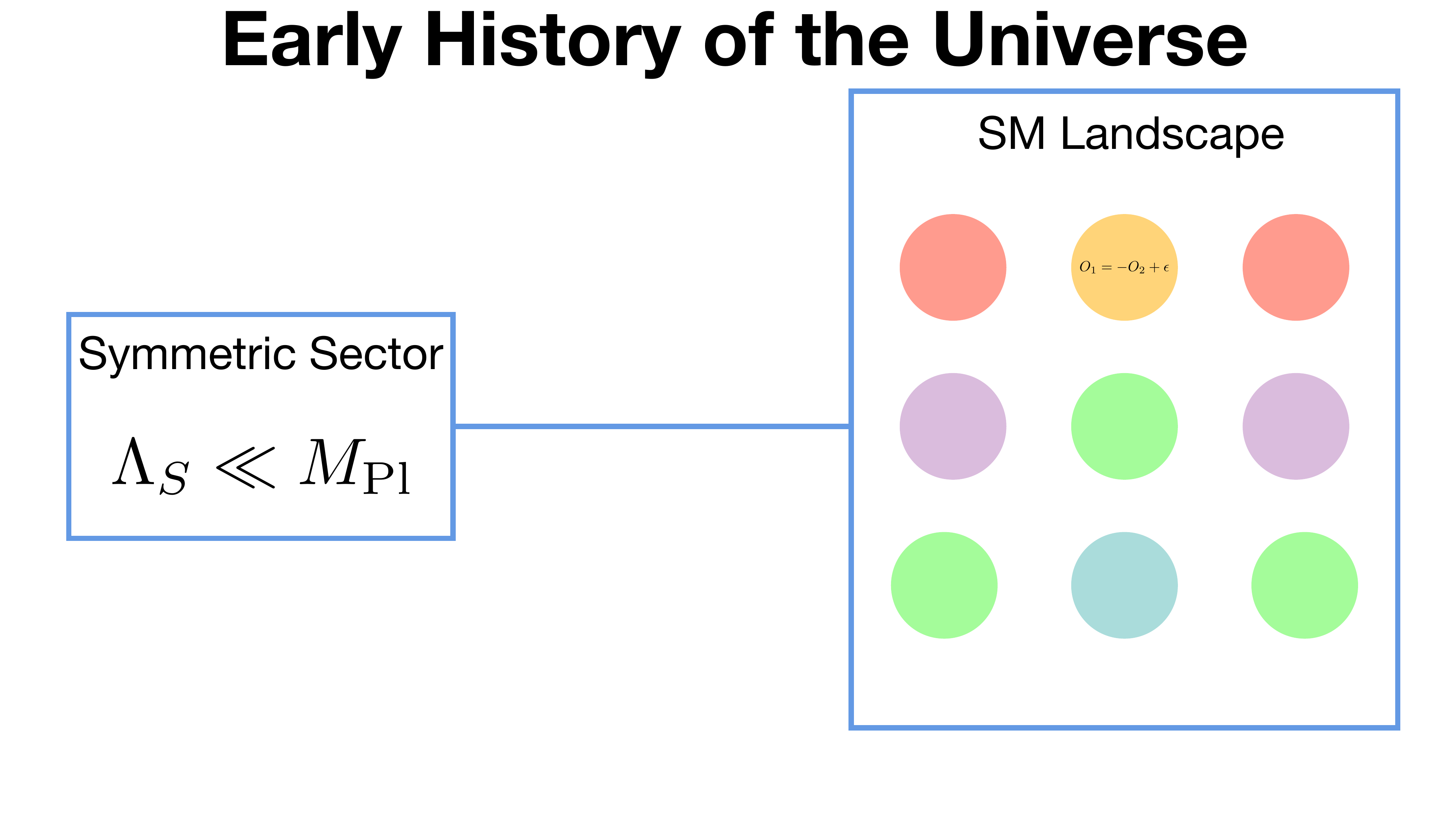} 
\includegraphics[width=0.45\textwidth]{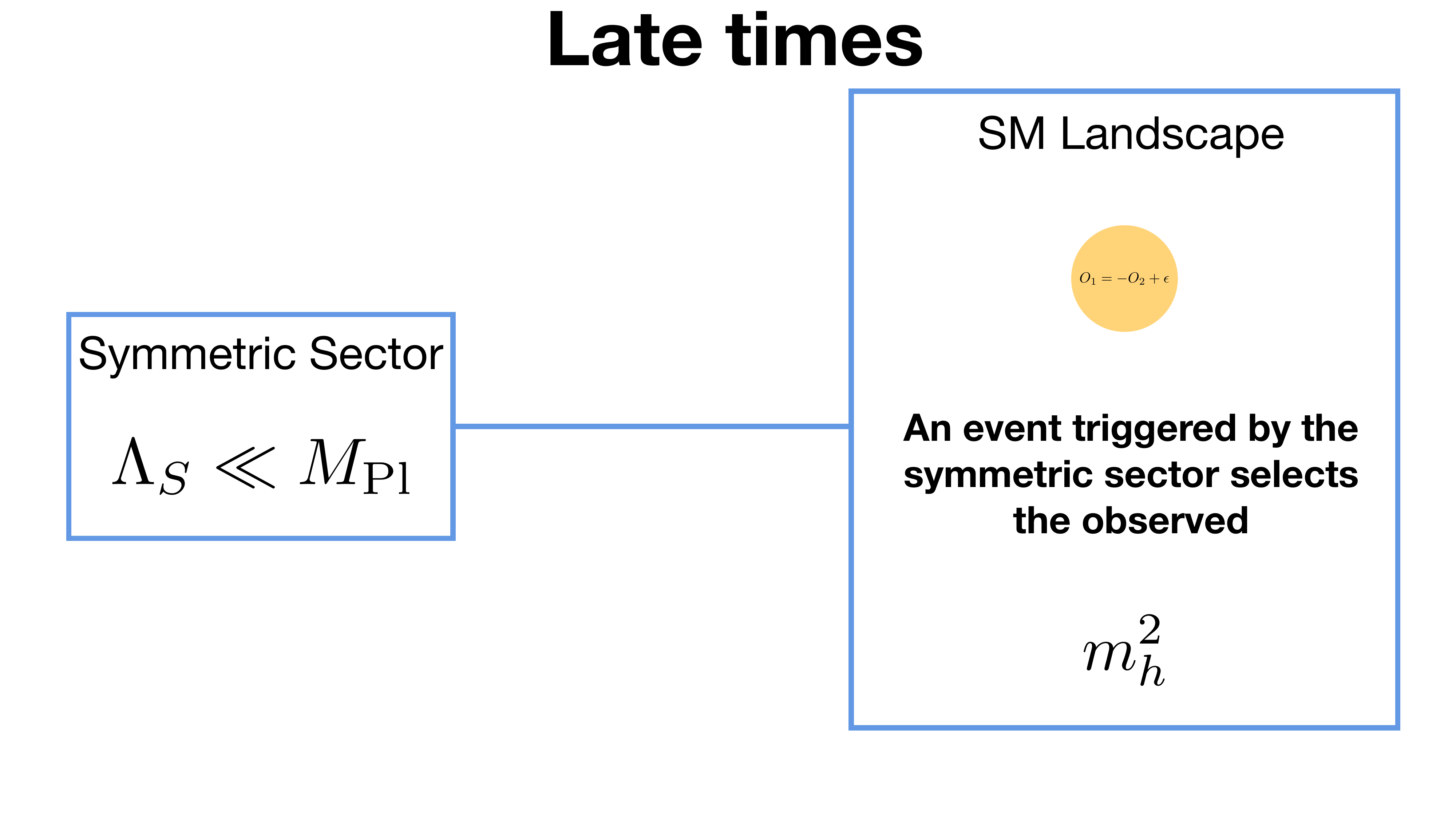} 
\caption{Models of cosmological selection of the weak scale. A symmetric sector, where a large hierarchy of scales is technically natural, is weakly coupled to a landscape of values of $m_h^2$. The SM landscape contains tuned values of $m_h^2$ including the observed one and is populated early in the history of the Universe. At a later time a cosmological event selects the observed value of $m_h^2$ through the coupling to the symmetric sector. Different selection mechanisms are shown in Fig.~\ref{fig:sketch2}.
\label{fig:sketch1}}
\end{figure}

Cosmological explanations of the weak scale have the schematic structure shown in Fig.~\ref{fig:sketch1}. Early in the history of the universe (left panel) we have a landscape of values for $m_h^2$ and a symmetric sector weakly coupled to the SM. In the symmetric sector a large hierarchy of scales is technically natural and it is not destabilized by the small coupling to the SM. The sector is {\it symmetric} in the sense that its approximate symmetries naturally stabilize a large hierarchy of scales. At late times, a cosmological event triggered by the Higgs vev and the coupling between SM and symmetric sector selects the observed value of the weak scale (right panel of Fig.~\ref{fig:sketch1}). 

The landscape of vacua can be realized in the form of causally disconnected patches of the Universe forming a Multiverse~\cite{Bousso:2000xa, Freivogel:2011eg, Winitzki:2008zz}, possibly populated during inflation. It is easy to always approximately decouple the landscape to the point of making detection prospects of the multitude of vacua almost non-existent. However it is useful to keep in mind that the more standard string theory (or field theory~\cite{ArkaniHamed:2005yv,Ghorbani:2019zic}) landscape is not the only option. The landscape can also be entirely contained in our patch of the Universe, either in the form of a scanning field coupled to the Higgs, as is the case for the Relaxion~\cite{Graham:2015cka}, or of feebly interacting copies of the Standard Model, as was proposed in Nnaturalness~\cite{Arkani-Hamed:2016rle}. 

\begin{figure}[t]
\includegraphics[width=0.45\textwidth]{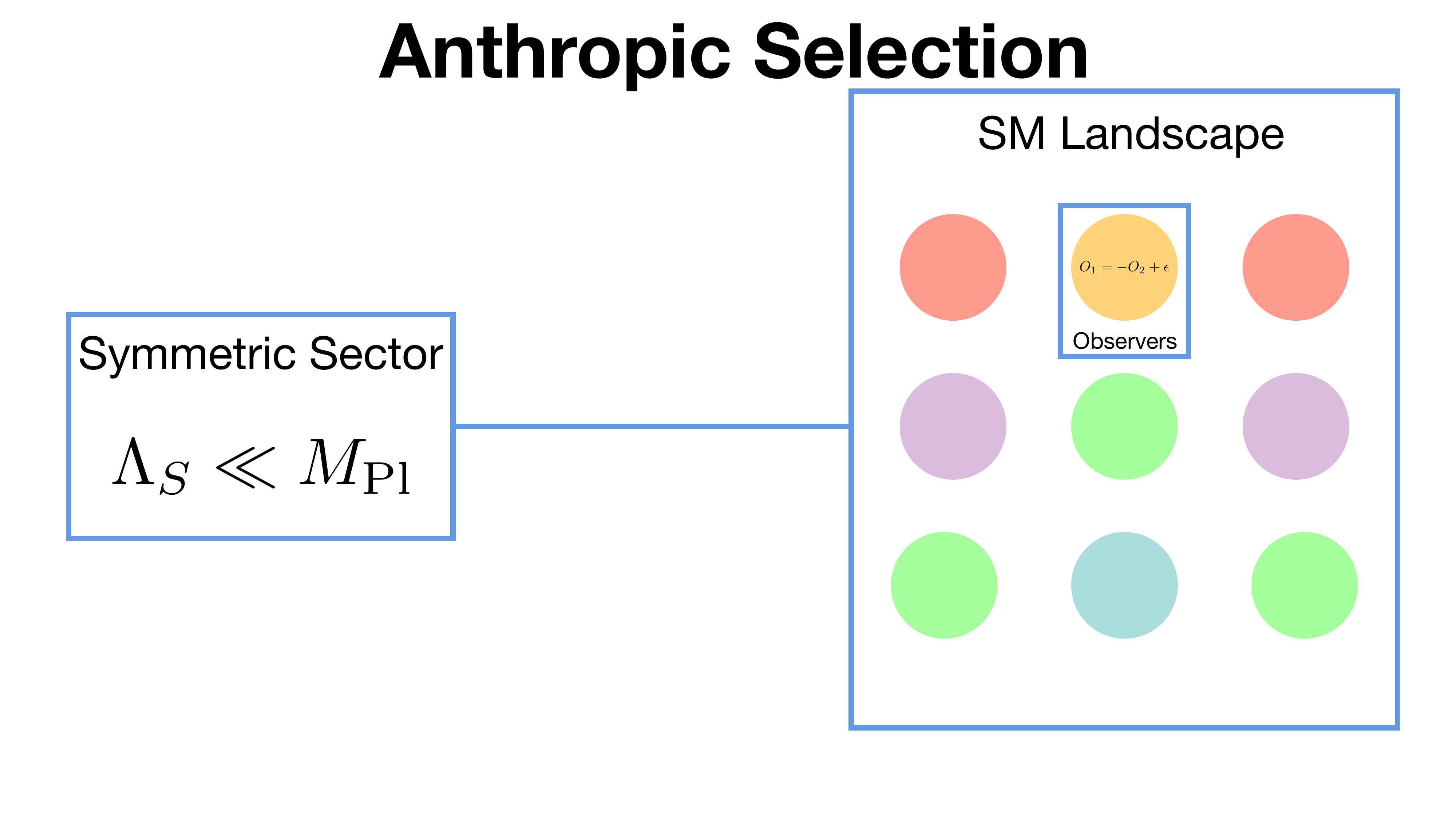} 
\includegraphics[width=0.45\textwidth]{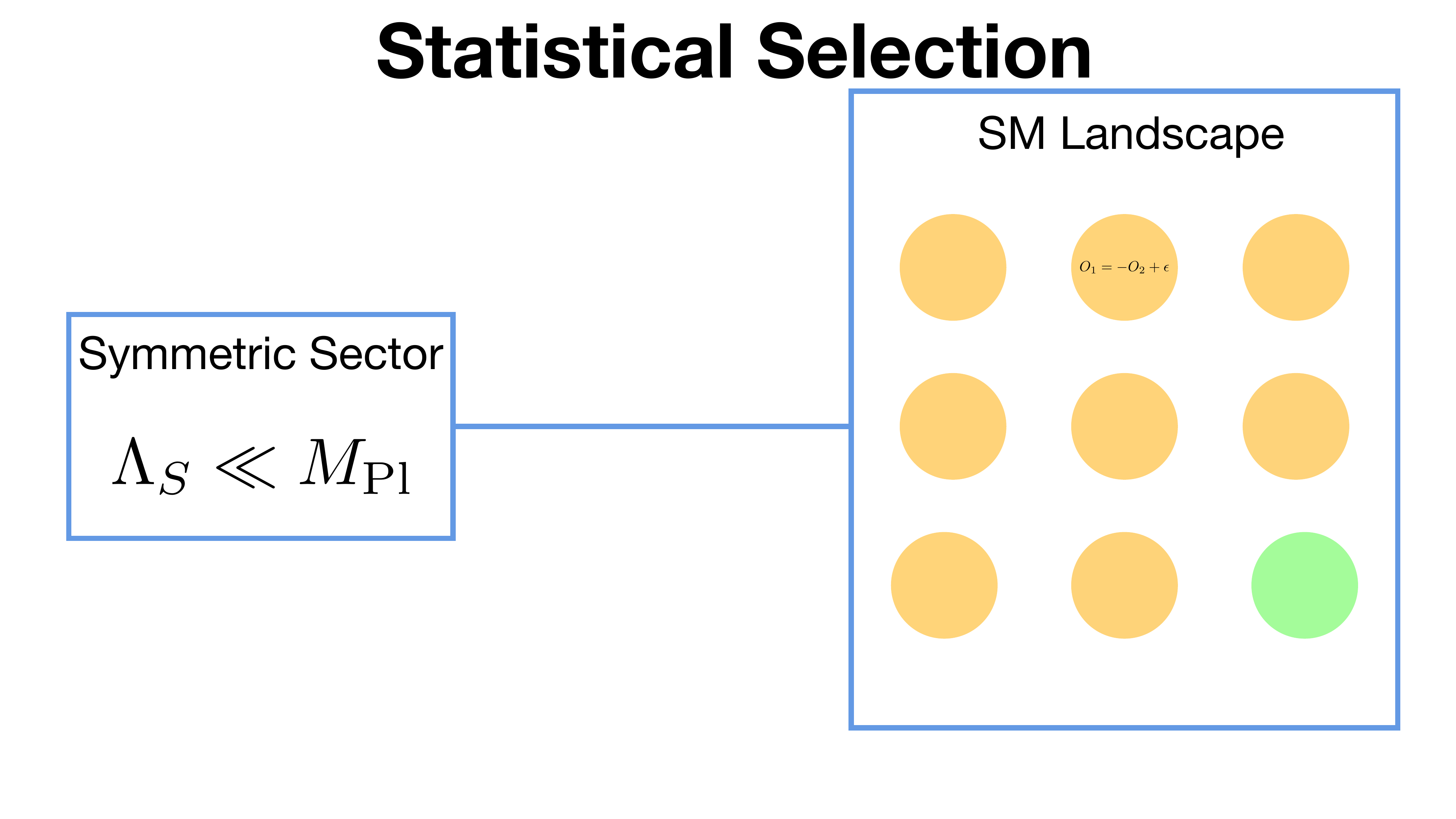} 
\includegraphics[width=0.45\textwidth]{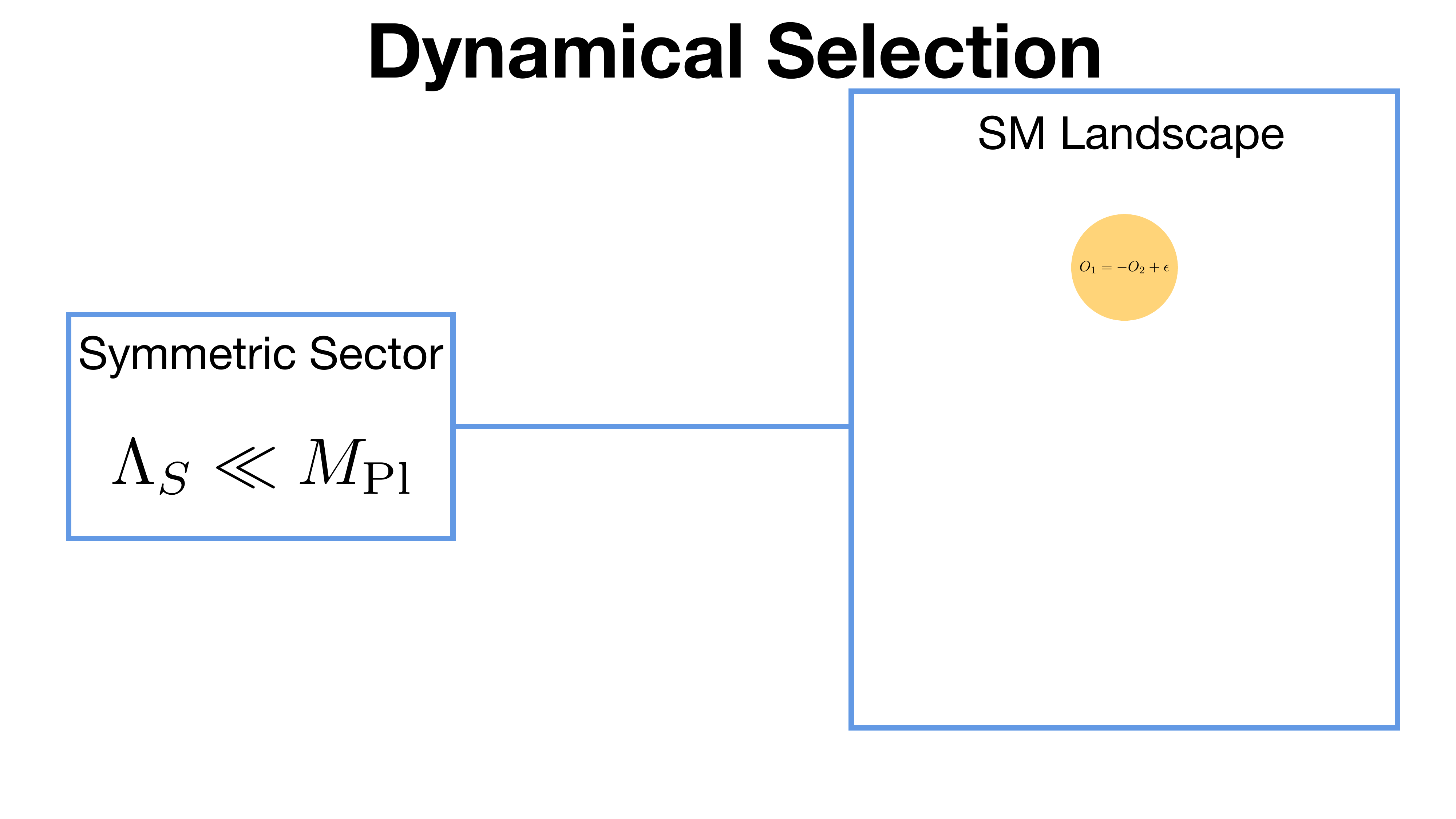} 
\caption{Models of cosmological selection of the weak scale. Anthropic selection (upper left panel), Statistical selection (upper right panel) and Dynamical selection (lower panel) are distinguished by the structure of the landscape at late times. In the anthropic case the landscape contains all values of $m_h^2$ with no preference for $\langle h \rangle \simeq v$. In the statistical case $\langle h \rangle \simeq v$ dominates Multiverse according to some measure, but also all other values are present. In the dynamical case only universes with $\langle h \rangle \simeq v$ are cosmologically long-lived and non-empty.
\label{fig:sketch2}}
\end{figure}

We identify three broad categories for the selection mechanism in Fig.~\ref{fig:sketch2}: 1) {\it Anthropic Selection}~\cite{Hall:2014dfa,DAmico:2019hih, Agrawal:1997gf, Arkani-Hamed:2004ymt}. Observers can arise only if $\langle h \rangle \simeq v$. 2) {\it Statistical Selection}~\cite{Dvali:2003br,Dvali:2004tma,Geller:2018xvz,Cheung:2018xnu,Giudice:2021viw}. Given some measure, the Multiverse is dominated by patches where $\langle h \rangle \simeq v$. 3) {\it Dynamical Selection}~\cite{Arvanitaki:2016xds,Arkani-Hamed:2020yna, Graham:2015cka, Arkani-Hamed:2016rle, Giudice:2019iwl, Strumia:2020bdy, Csaki:2020zqz, TitoDAgnolo:2021nhd}. Only non-empty\footnote{The simplest definition of an empty a patch is given by a universe where a positive CC always dominates the energy density. However for our purposes it is sufficient that, as explained below, observers can only exist for a sufficiently short time. We can consider empty also patches where the CC is positive and larger than a certain threshold $\Lambda > \Lambda_{\rm min}$. In these patches we can have a period of radiation and/or matter domination that lasts at most $\sim M_{\rm Pl}/\sqrt{\Lambda_{\rm min}}$. For an empty patch this time has to be much shorter compared to the age of our universe. In most models this time is much shorter than typical particle physics scales ($\ll 1/v$).} patches where $\langle h \rangle \simeq v$ live for cosmologically long times. 

Anthropic and statistical selection do not require new observable physics coupled to the SM. The mechanism that populates the landscape and generate its structure can take place at unobservably high energies or be due to non-dynamical fields with extremely feeble couplings to the SM~\cite{Dvali:2003br, Dvali:2004tma, Giudice:2019iwl, Giudice:2021viw}. 

Dynamical selection occurs when at early times we have a ``standard" landscape, with no preference for small $\langle h \rangle$, but at late times only universes with $\langle h \rangle \simeq v$ exist and are not empty. The distinction between this class of ideas and anthropic selection might seem blurred. However there are one conceptual difference and one (more important) practical difference. The conceptual difference is that dynamical selection mechanisms do not require the absence of observers from other patches of the Multiverse. The ``wrong" values of the Higgs vev are matter and/or radiation dominated for a very short time compared to the age of the observable Universe (often even compared to particle physics scales). During this time, an observer whose typical timescales are $1/M \ll 1/v$ can possibly exist, but it does not change the statement that the only way to have a universe even remotely resembling our own is to have $\langle h \rangle \simeq v$. The practical difference is that dynamical selection requires new physics coupled to the Higgs and can be detected in the near future. From now on we focus on this class of models that does not suffer from measure problems and has the best chance of being tested experimentally. Our idea belongs to this category.

Having said this, it is clear that what we have called dynamical solutions have anthropic elements. First of all, most of them, including our proposal, rely on Weinberg's argument to explain the CC. Secondly, the existence of a macroscopic, long-lived and non-empty universe \emph{is} Weinberg's argument. We have already argued that dynamical solutions, unlike anthropic ones, do not require the absence of observers from other universes, but we can see how this conceptual point can be the starting point of endless debates. However we find that the distinction between these two classes of ideas has practical value in light of the important phenomenological distinctions that we now discuss.

In existing ``dynamical" models the selection mechanism is composed of two ingredients: 1) one or more new scalars or pseudo-scalars with masses inversely proportional to the cutoff of the Higgs sector and 2) an operator whose vev is a monotonic function of the Higgs vev. These operators are coupled to the new scalar(s) and were collectively identified as {\it triggers} in~\cite{Arkani-Hamed:2020yna}. When the Higgs vev (and thus the operator vev) crosses certain upper or lower bounds, a cosmological event is triggered via the coupling to the new scalar(s). 

In the next two Subsections we show why we expect new particles with masses inversely proportional to the cutoff and how the choice of trigger operator determines the phenomenology of dynamical selection. Our considerations apply to the majority of these models, but exceptions to the power counting arguments in the next Section exist, either because the weak scale is not selected by comparing two different terms in the potential of a new scalar, but rather directly its mass to that of SM particles~\cite{Arkani-Hamed:2016rle} or because it occurs via a non-dynamical field~\cite{Giudice:2019iwl}.

\subsection{Cosmological Naturalness Power Counting}

\begin{figure}[t]
\begin{align*}
&\parbox{8em}{\includegraphics[width=8em]{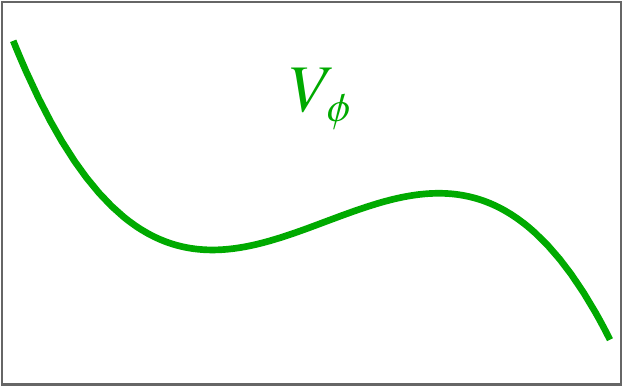} }\qquad &+& \qquad &\parbox{8em}{\includegraphics[width=8em]{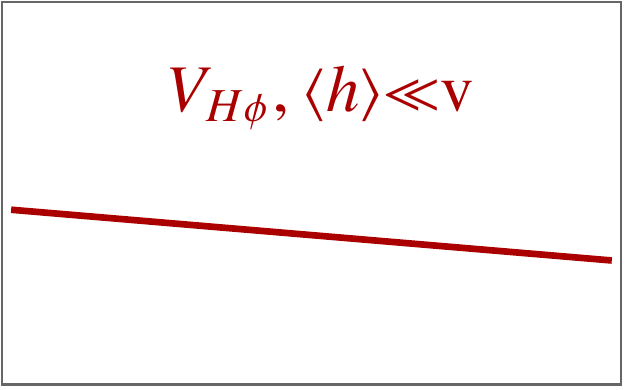}}  \qquad &=& \qquad &\parbox{8em}{\includegraphics[width=8em]{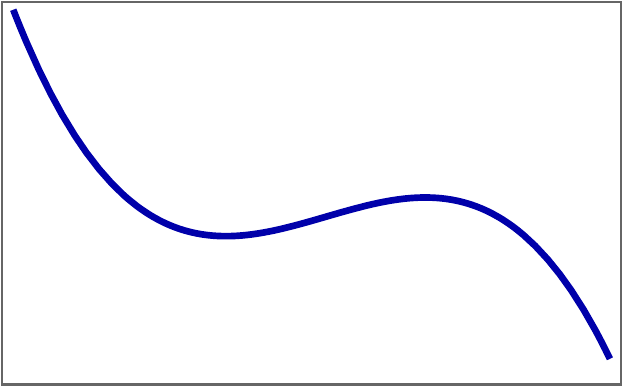}}  \\[1em]
&\parbox{8em}{\includegraphics[width=8em]{L} }\qquad &+& \qquad &\parbox{8em}{\includegraphics[width=8em]{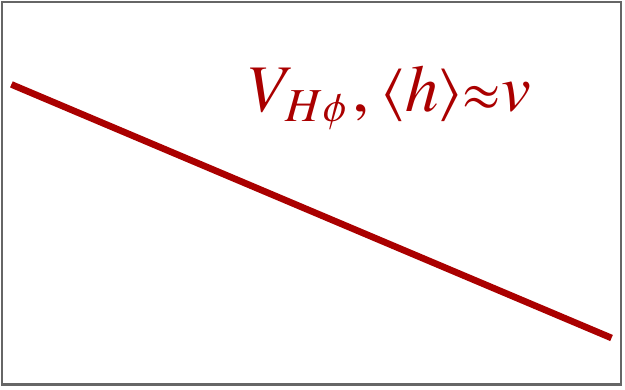}}  \qquad &=& \qquad &\parbox{8em}{\includegraphics[width=8em]{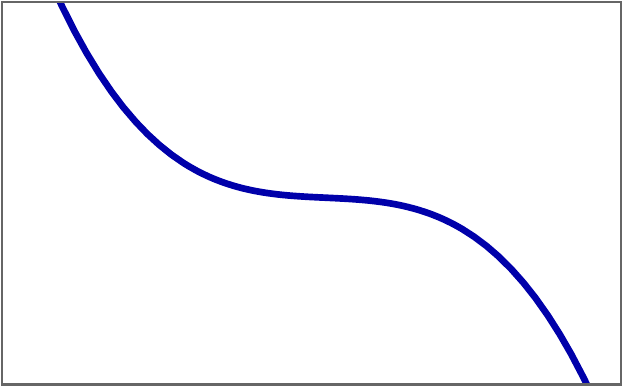}}  \\[1em]
&\parbox{8em}{\includegraphics[width=8em]{L} }\qquad &+& \qquad &\parbox{8em}{\includegraphics[width=8em]{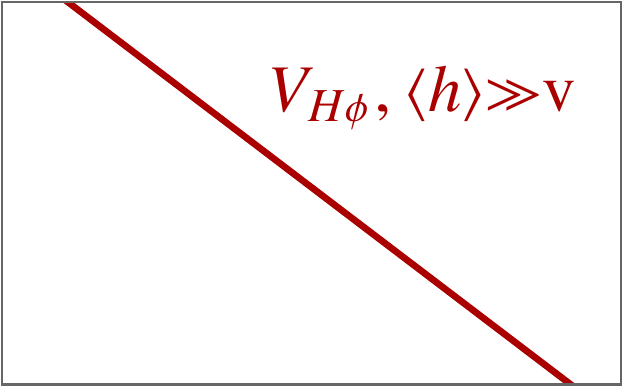}}  \qquad &=& \qquad &\parbox{8em}{\includegraphics[width=8em]{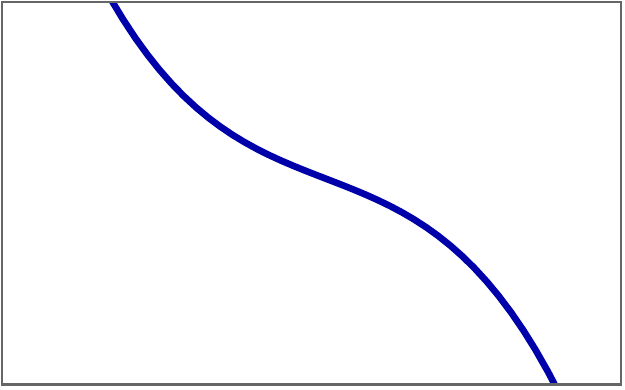}}  \\
\end{align*}
\caption{Schematic structure of how the Higgs-dependent potential $V_{H \phi}$ can affect the scalar potential $V_\phi$ to trigger a qualitative change when $\langle h \rangle \simeq v$.
\label{fig:NDA}}
\end{figure}

The presence of new light scalars $\phi$, in many of the models that dynamically select the weak scale in the early history of the Universe, can be understood from a simple parametric argument. Neglecting $\mathcal{O}(1)$ factors we can write any term in the $\phi$ potential as
\be
V_\phi \supset m_\phi^2 M_*^2 \left(\frac{\phi}{M_*}\right)^m \, .
\ee
Here and in the following, we restore units of $\hbar$~\cite{Georgi:1992dw} to infer the correct parametrics. However, for simplicity, we keep giving formulas in natural units $\hbar = 1$. If $\hbar \neq 1$ masses and scalar fields/vevs have different dimensions and we will be careful about this distinction. In our formulas $M_*$ is a cutoff scale (with the same dimensions as $\phi$), whereas $m_\phi$ is a mass. Dimensionally, $\text{mass} = \text{coupling} \times \text{scale}$.

We can now include an interaction between $\phi$ and the Higgs boson. We denote the cutoff scale of the Higgs sector by $\Lambda_H$ and by $\tilde v \leq v$ possible light SM or BSM scales, not depending explicitly on the Higgs vev $\langle h \rangle$. Then, integrating out the SM {\it at tree-level} we have
\be
V_{\langle H\rangle\phi} \simeq \mu^2 M_*^2 \left(\frac{\phi}{M_*}\right)^n \frac{\tilde v^{2q-j}\langle h \rangle^j}{\Lambda_H^{2q}}\, ,
\ee
with $q\geq 1$ and $j>0$. Examples of couplings of $\phi$ to the SM present in the literature include: 1) $\phi {\rm Tr}[G\widetilde G]$~\cite{Dvali:2003br,Dvali:2004tma,Graham:2015cka,Geller:2018xvz,TitoDAgnolo:2021nhd}, giving
\be
\tilde v^{2q-j}\langle h \rangle^j \simeq f_\pi^3\langle h \rangle 
\ee
for QCD (note that $f_\pi$ depends on $\langle h \rangle$). A similar result holds for BSM gauge groups whose quarks get part of their mass from the Higgs. 2) $\phi^n H_1 H_2$~\cite{Arkani-Hamed:2020yna}: 
\be
\tilde v^{2q-j}\langle h \rangle^j = \frac{\sin2\beta}{2} \langle h \rangle^2
\ee
and 3) $\phi^n |H|^2$~\cite{Cheung:2018xnu,Strumia:2020bdy, Csaki:2020zqz}: $\tilde v^{2q-j}\langle h \rangle^j=\langle h \rangle^2$. 

To select the weak scale, we need the Higgs-induced part of the potential $V_{\langle H \rangle \phi}$ to be comparable to the Higgs-independent part $V_\phi$ when $\langle h \rangle \simeq v$, as sketched in Fig.~\ref{fig:NDA}. Alternatively, if the mechanism involves, for instance, stopping a slow-rolling scalar, we want the first derivatives with respect to $\phi$ to be comparable~\cite{Graham:2015cka}. With our parametrization of the potential these two conditions lead parametrically to the same result 
\be
\frac{m_\phi^2}{\mu^2} \simeq \frac{\tilde v^{2q-j}v^j}{\Lambda_H^{2q}}\lesssim \frac{v^{2q}}{\Lambda_H^{2q}}\, . \label{eq:par1}
\ee
This shows that the separation between the weak scale and the Higgs cutoff is given by an approximate symmetry on $\phi$  that protects its mass and potential. Furthermore, it gives a smoking-gun signature for these models. If we measure the $\phi$ mass, its coupling to the SM $\mu$ and the Higgs cutoff $\Lambda_H$, we can test Eq.~\eqref{eq:par1}.

We can go even further and  obtain an upper bound on $m_\phi$ that depends only on the cutoff scales $M_*$, $\Lambda_H$, by noticing that $\mu^2$ has two upper bounds. One is determined by experiment, since $\mu^2$ sets the strength of $\phi$ interactions with the SM. The other one comes from quantum corrections, since integrating out the SM beyond tree-level can generate contributions to $V_\phi$, but to select the weak scale $V_\phi$ cannot be too large (i.e. it has to be comparable to the tree-level Higgs-induced potential $V_{\langle H \rangle \phi}$ when $\langle h \rangle \simeq v$). We now use these constraints to derive upper bounds on $m_\phi$ for three different types of couplings of $\phi$ to the SM.

\begin{figure}[t]
$$\parbox{0.4\textwidth}{\includegraphics[width=0.4\textwidth]{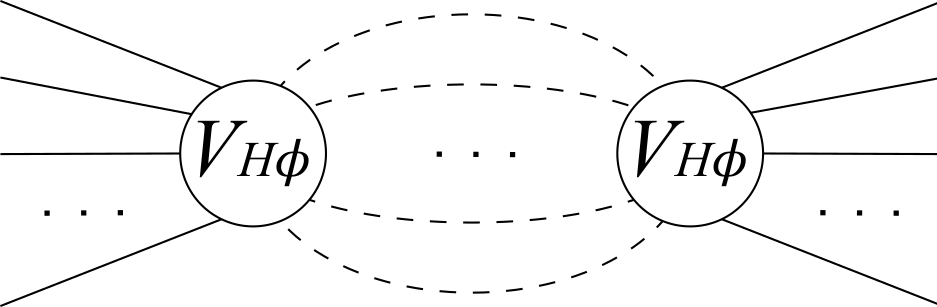}} \qquad \sim \qquad \left(\frac{\mu^2 M_*^2}{\Lambda_H^{2q}}\right)^2 \times \frac{(g_H \Lambda_H)^{4(2q - 1) - 2 \cdot 2q}}{(16 \pi^2)^{2q - 1}}$$
\caption{Schematic diagram giving loop corrections to the potential $V_\phi$ from two insertions of $V_{H\phi}$. The scalar $\phi$ is denoted by a continuous line, the $2 q$ Higgs propagators by dashed lines. \label{fig:insertions}}
\end{figure}

The simplest example is given by the $\phi |H|^2$  coupling. Let us first consider the impact of quantum corrections on $\mu$. In this case the leading contribution to $V_\phi$ is from a single insertion of $V_{H\phi}$,
\be
V_{H\phi}= \mu^2 M_* \, \phi \, \frac{|H|^2}{\Lambda_H^2}\, .
\ee
By closing the Higgs loop we see that (barring fine-tuning) $m_\phi^2 \gtrsim g_H^2 \mu^2/16 \pi^2$, with $g_H$ being a coupling in the Higgs sector. This takes into account that Higgs loop integrals are cut off by a mass scale $g_H \Lambda_H$ (and not a vev $\Lambda_H$).
Eq.~\eqref{eq:par1} supplemented by this condition on $\mu$ shows that a cosmological selection mechanism with the trilinear coupling $\phi |H|^2$ can solve only the little hierarchy problem
\be
g_H \mu \lesssim 4 \pi m_\phi \rightarrow g_H \Lambda_H \lesssim 4\pi v\, .
\ee
To get the bound on $m_\phi$ we may use the fact that $\mu$ has an experimental bound $\mu < \mu_{\rm exp}(m_\phi;M_*,\Lambda_H)$, so that \eqref{eq:par1} gives:
$m_\phi \lesssim \mu_{\rm exp}$. We do not give the explicit value of $\mu_{\rm exp}$ since it depends strongly on $m_\phi$. In Fig.~\ref{fig:DM} we plot it in terms of $\kappa \equiv \mu^2 M_*/(\Lambda_H^2 m_\phi)$ for $m_\phi \lesssim \textrm{eV}$.

Instead, if the leading contribution to $V_{\phi}$ arises from two (or more) insertions of $V_{H\phi}$ (for instance in the $\phi H_1 H_2$ case) we have 
\be
\frac{g^{4q-4}_H}{(16 \pi^2)^{2 q - 1}} \frac{\mu^4 M_*^4}{\Lambda_H^4} \lesssim m_\phi^2 M_*^2 \, ,
\ee
as shown in Fig.~\ref{fig:insertions}, assuming for simplicity $j=2q$, so that extra light scales $\tilde{v}$ are absent. If we put this together with Eq.~\eqref{eq:par1} we obtain 
\be
m_\phi \lesssim \frac{g_H^2\Lambda_H^2}{4 \pi M_*} \left( \frac{4 \pi v}{g _H \Lambda_H}\right)^{2q}\lesssim \frac{4 \pi v^2}{M_*}\, ,
\ee
where the last inequality is valid in the $\phi H_1 H_2$ case, i.e. $2q = 2$.  We can raise the cutoff all the way to $M_{\rm Pl}$, predicting very light scalars with $m_\phi \lesssim v^2/M_*$. 

As our last example, we consider the coupling $(\phi/{M_*}){\rm Tr}[G\widetilde G]$. With this choice, quantum corrections do not give us any information on $m_\phi$ beyond Eq.~\eqref{eq:par1}. In this case experiment is more useful. Stringent bounds on axion couplings allow us to conclude 
\be
m_\phi^2 \lesssim \mu^2_{\rm exp} \frac{v f_\pi^3}{\Lambda_H^4} \simeq (0.1\; {\rm eV})^2 \left(\frac{10^8{\rm GeV}}{{M_*}}\right)^2
\ee
for QCD. A similar discussion holds for $(\phi/{M_*}){\rm Tr}[F\widetilde F]$ with a new non-abelian gauge group whose charged fermions have a $\langle h \rangle$-dependent mass~\cite{Graham:2015cka}. 

These three examples make more precise the intuition from Eq.~\eqref{eq:par1}. The separation between the Higgs vev and the cutoff is made stable by a symmetry protecting $m_\phi$. They also provide a second type of inequalities that can be used to test these mechanisms: the bigger the cutoff $M_*$ of the $\phi$ sector the lighter we expect the new scalars to be. Note that Eq.~\eqref{eq:par1} on its own, in the $\phi {\rm Tr}[G\widetilde G]$ case, does note give an experimentally interesting relation between $m_\phi$ and $\Lambda_H$, because $\mu$ depends on $\Lambda_H$ in a way that cancels it from the equation.

To conclude we remark that one can couple $\phi$ to the SM more weakly than what naturalness or experiment require, making it even lighter. The dilaton in~\cite{Csaki:2020zqz}, $m_\chi \simeq {\rm MeV}-{\rm GeV}$, saturates our upper bound for the cutoff in the paper ${M_*\simeq}\;\Lambda_H\simeq$~few TeV. On the contrary, the scalar in~\cite{Strumia:2020bdy} is much lighter $m_\phi \lesssim v^4/M_{\rm Pl}^3$ even if the same $|H|^2$ trigger was used and the cutoff is of a similar order. The relation in Eq.~\eqref{eq:par1} between the $\phi$ mass and the coupling to the SM remains valid. This gives an interesting target to laboratory searches, as we discuss in Section~\ref{sec:DM} in the context of dark matter.

\subsection{Trigger Operators and Low Energy Predictions}

The second generic prediction of mechanisms selecting the weak scale dynamically is old or new physics with relatively small mass $m \lesssim 4\pi m_h$ coupled at $\mathcal{O}(1)$ to the Higgs. This is what we have called the trigger, i.e. the local operator whose vev depends on $\langle h \rangle$. We have already seen in the previous Section that four examples exist in the literature: $\phi {\rm Tr}[G\widetilde G], \phi H_1 H_2, \phi |H|^2, \phi {\rm Tr}[F\widetilde F]$, where $G$ is the QCD field strength and $F$ the field strength of a BSM gauge group. Clearly the choice of trigger is central to the phenomenology of the model. From the point of view of experiment, models of cosmological naturalness can be conveniently classified based on their trigger. For example, theories with a $\phi {\rm Tr}[G\widetilde G]$ coupling predict axion-like phenomenology at low energy, while theories with $\phi H_1 H_2$, Equivalence-Principle-violating light scalars and a new Higgs doublet.

In the SM we essentially have only one possible category of operators that can act as a trigger, given by divergences of non-gauge invariant currents: ${\rm Tr}[G\widetilde G]$ and ${\rm Tr}[W\widetilde W]$. In this case QCD and EW interactions are the physics coupled to the Higgs, characterized by mass scales comparable or smaller than $m_h$. However purely within the SM the weak $\theta$-angle is not observable~\cite{Shifman:2017lkj}.

Constructing BSM triggers requires introducing new physics coupled to the Higgs. For instance we can have a second Higgs doublet and the operator $\mathcal{O}_T=H_1 H_2$~\cite{Dvali:2001sm,  Arkani-Hamed:2020yna,Espinosa:2015eda} or a new confining gauge group whose fermions have a Yukawa coupling to the Higgs~\cite{Graham:2015cka} with trigger operator $\mathcal{O}_T={\rm Tr}[F\widetilde F]$.
In general if we introduce in the BSM theory masses much larger than $m_h$ the vev of the trigger operators will be proportional to those scales rather than $v$, just from dimensional analysis. This is one of the familiar incarnations of the hierarchy problem, i.e. dimensional analysis works.

Other examples of triggers that might work in extensions of the SM are ${\rm Tr}[W\widetilde W]$ or higher dimensional operators breaking baryon and/or lepton number. Both options require adding to the SM new baryon and/or lepton number breaking sensitive to the Higgs vev. To assess the feasibility of these ideas a phenomenological study comparable in scope to the one performed in~\cite{Arkani-Hamed:2020yna} for $H_1 H_2$ is needed.

The difficulty in finding BSM ``trigger" operators $\mathcal{O}_T$ lies in the requirement that $\langle \mathcal{O}_T\rangle$ must be sensitive to the Higgs vev. In general we need new particles coupled at $\mathcal{O}(1)$ to the Higgs whose typical mass scales are at most comparable to the weak scale. Beyond the SM it is extremely challenging to find new physics with these characteristics not already excluded by the LHC. Currently viable models, as the type-0 2HDM proposed in~\cite{Arkani-Hamed:2020yna}, which leads to the operator in~\eqref{eq:H1H2basic}, are on the verge of being discovered or excluded. A similar phenomenological analysis has been performed for ${\rm Tr}[F\widetilde F]$ in~\cite{Beauchesne:2017ukw}. 

In practice only a limited number of trigger operators is viable and each trigger can be used in many different ways to select the Higgs mass. For example ${\rm Tr}[G\widetilde G]$ is used in~\cite{Dvali:2003br,Dvali:2004tma,Graham:2015cka,Geller:2018xvz,TitoDAgnolo:2021nhd}. So each trigger identifies phenomenology that is generically associated to Higgs naturalness, independently of a specific construction. 

This feature is generic to a large class of models that select the observed value of the weak scale in the early history of the Universe: only a few choices of couplings to the SM are possible. This leads to unified expectations for their phenomenology and the concrete possibility of testing in the near future the concept of cosmological naturalness for the Higgs mass. 

\section{Description of the Mechanism} \label{sec:mechanism}
After this preliminary discussion, we introduce our mechanism to select the electroweak scale.

\subsection{Basic Idea}
\begin{figure}[t]
$$\includegraphics[height=10em]{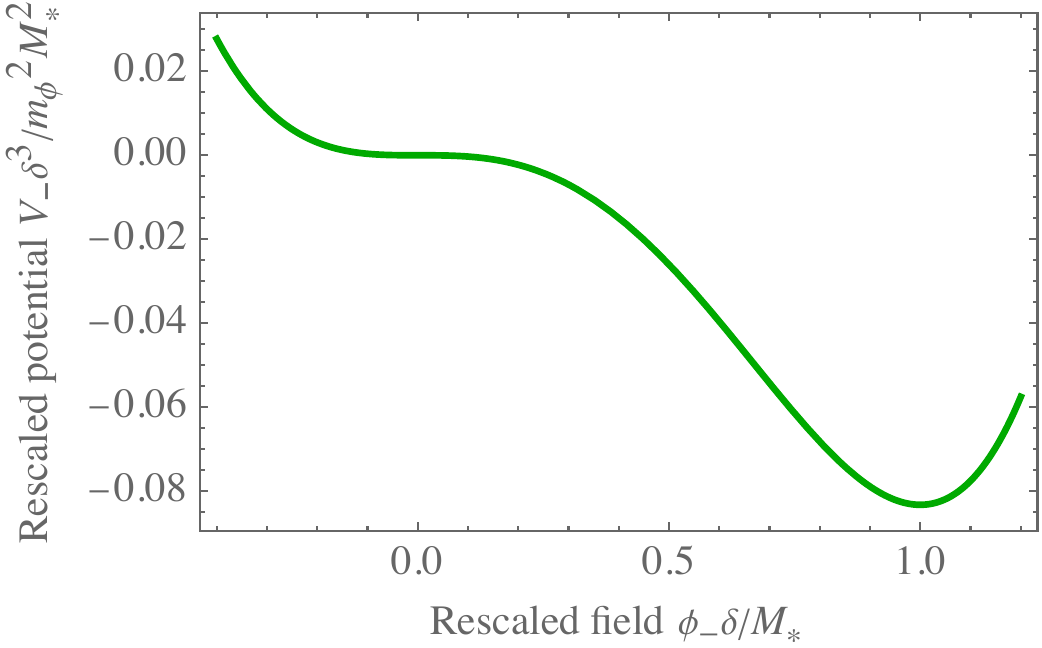} \qquad \includegraphics[height=10em]{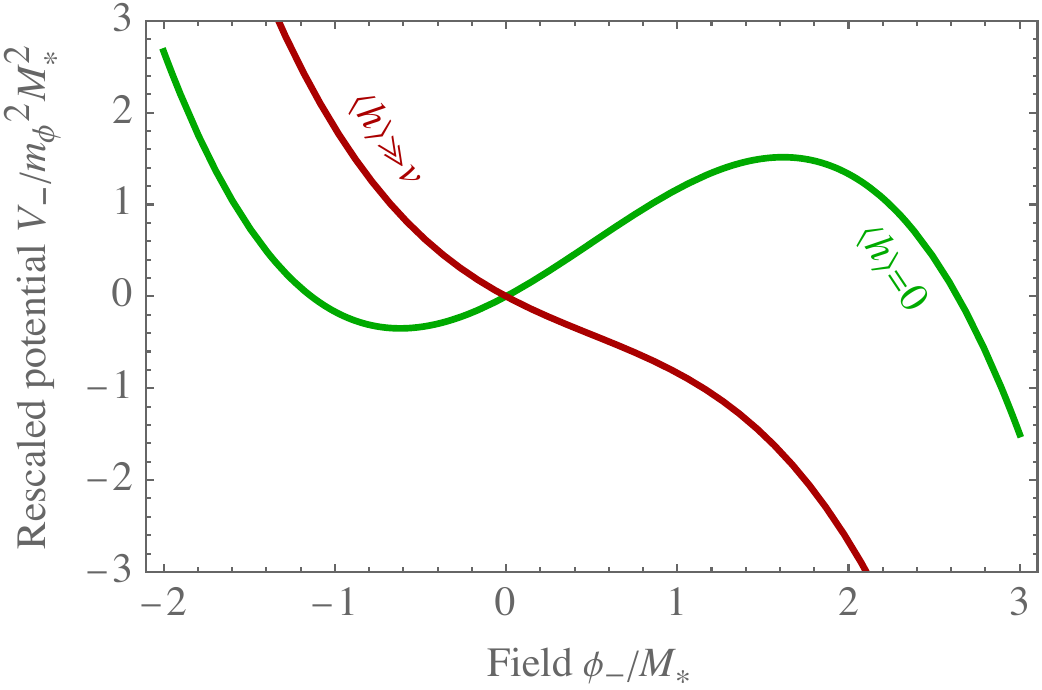} $$
\caption{Example potential $V_-(\phi_-)$ with two widely separated minima. The right panel zooms in close to the safe local minimum at $\phi \sim M_*$. This is destabilised if the Higgs acquires a large vev (red line). Note the different rescaling in the two panels for both the field and the potential.  \label{fig:potential}}
\end{figure}

At low energy the theory includes a new scalar $\phi_-$ with an approximate shift symmetry. The $\phi_-$ potential has two widely separated minima. The deepest minimum of the potential has energy density of $\mathcal{O}(-M^2 M^2_*)$ with $M$ the largest mass scale in the theory and $M_*= M/g_* \sim M$ a vev associated to it. This energy density is $\mathcal{O}(1)$ larger than the largest cosmological constant in the landscape. Universes where $\phi_-$ rolls to this minimum rapidly crunch. The shallow ``safe" minimum of $\phi_-$ has energy density $\mathcal{O}(m_{\phi_-}^2 M_*^2)$, with $m_{\phi_-} \ll M$. In this minimum the CC can be scanned finely around zero. Its observed value today can, for instance, be selected by Weinberg's anthropic argument~\cite{Weinberg:1988cp}. The $\phi_-$ potential is schematically depicted in the left panel of Fig.~\ref{fig:potential}. 

\begin{figure}[t]
$$\includegraphics[height=10em]{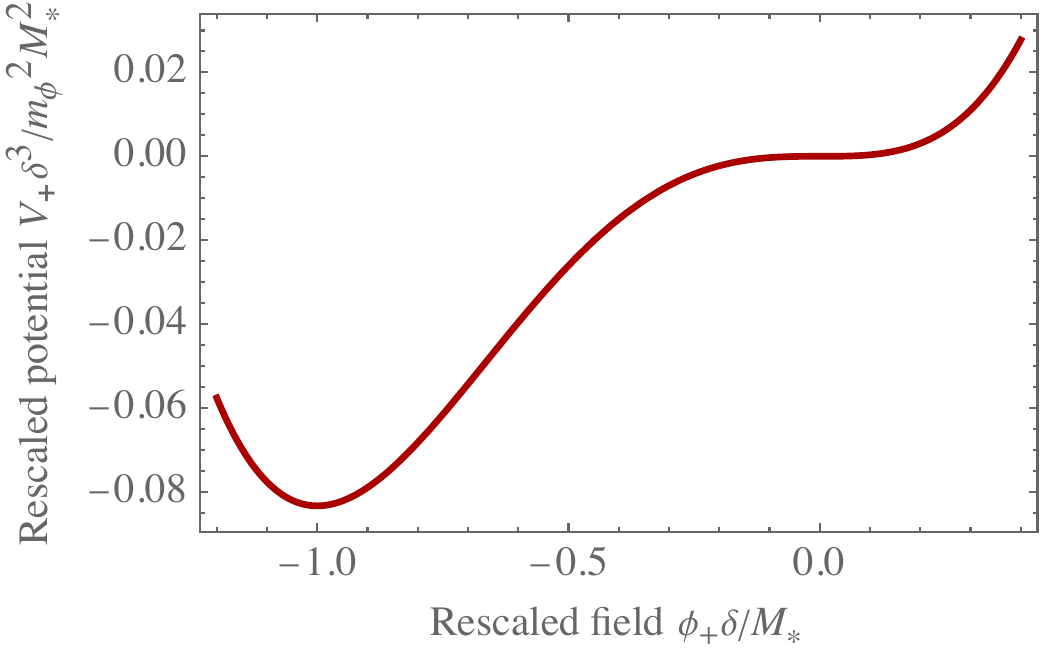} \qquad \includegraphics[height=10em]{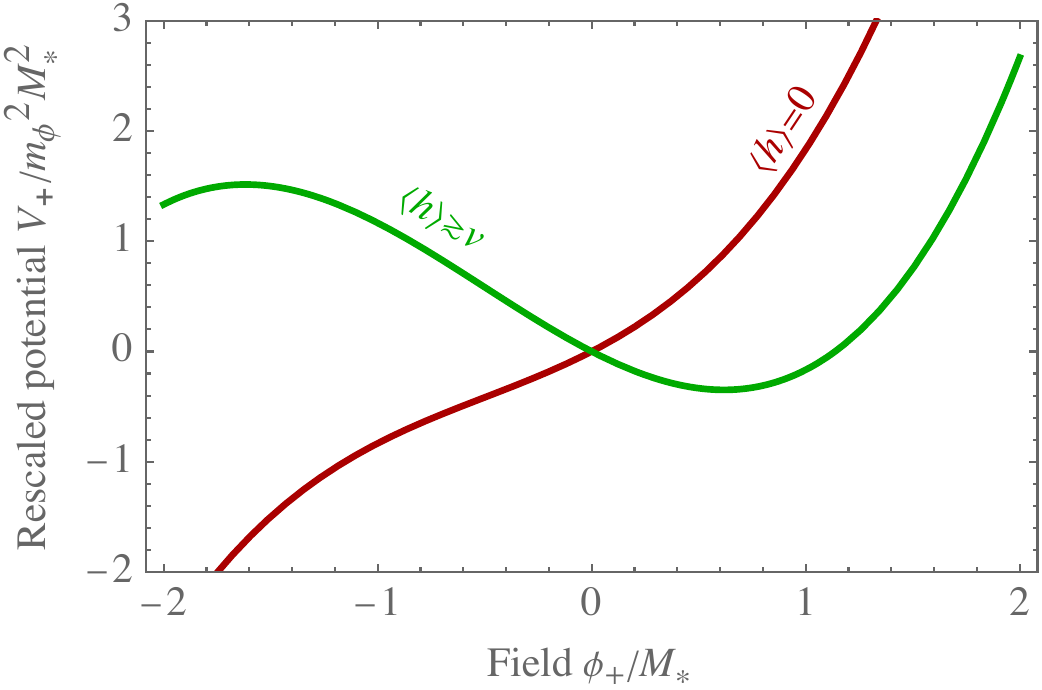} $$
\caption{Example potential $V_+(\phi_+)$, that selects a nonzero Higgs vev. The right panel zooms in close to the safe local minimum at $\phi_+ \sim M_*$, present only if the Higgs acquires a sufficiently large vev (green line). \label{fig:potential_phip}}
\end{figure}

A small value for the Higgs vev, $\langle h \rangle \ll M_*$, is selected by a $\langle h \rangle$-dependent tadpole in the $\phi_-$ potential. This tadpole destabilizes the safe metastable minimum when the Higgs vev is larger than $v$. The tadpole is generated by a coupling of $\phi_-$ to an operator $\mathcal{O}_T$ 
\be
V_{H \phi}= - a \phi_- \mathcal{O}_T +{\rm h.c.} \label{eq:tadpole}
\ee
whose vev is a monotonic function of $\langle h \rangle$. When $\langle h \rangle \gg v$ the tadpole in Eq.~\eqref{eq:tadpole} dominates the $\phi_-$ potential around $M_*$ and destroys the safe minimum (see Fig.~\ref{fig:potential}), so all universes with large and negative Higgs mass squared rapidly crunch. The small number that separates the weak scale from the cutoff $M_*$ is $m_\phi$, i.e. universes where the tadpole dominates near the metastable minimum of $\phi_-$,
\be
\frac{a \langle\mathcal{O}_T \rangle}{m_\phi^2 M_*} \gg 1, 
\ee
 are those which crunch fast.  The separation between $m_{\phi_-}$ and $M_*$ is technically natural, because $\phi_-$ is part of a very weakly coupled sector that can naturally be approximately scale-invariant or supersymmetric, without any measurable trace of scale invariance or supersymmetry in the SM. 
 
 The basic ``crunching" setup is conceptually the same as~\cite{Strumia:2020bdy,Csaki:2020zqz}, but, as we will see in more detail in the following, there are two important differences: 1) differently from \cite{Csaki:2020zqz} in our case inflation can happen at a very high scale and possibly be eternal. Crunching of patches where $\langle h \rangle \gg v$ occurs after reheating at temperatures below $v$, independently of the details of inflation. 2) In~\cite{Csaki:2020zqz} the SM becomes approximately scale invariant already above a few TeV. In~\cite{Strumia:2020bdy} new physics that protects the Higgs mass must appear at a few TeVs. Here and in our companion paper \cite{TitoDAgnolo:2021nhd} the symmetries protecting the $\phi_-$ potential can be invisible in the SM sector.

We have seen that $\phi_-$ stabilizes a hierarchy between the Higgs vev and the cutoff, but we can still have universes with vanishing Higgs vev. Universes with small (or vanishing) Higgs vevs are destabilized by an additional scalar $\phi_+$ coupled to $\mathcal{O}_T$ in the same way as $\phi_-$. The main difference is that $\phi_+$ does not have a safe metastable minimum when $\langle h \rangle =0$. This minimum is generated only if $\langle h \rangle \gtrsim v$. Then, as shown in Fig.~\ref{fig:potential_phip}, the universe rapidly crunches unless the Higgs acquires a sufficiently large vev. The mechanism with both scalars $\phi_\pm$ selects a \textit{small} and \textit{non-zero} Higgs vev.

\begin{figure}[t]
\includegraphics[width=0.5\textwidth]{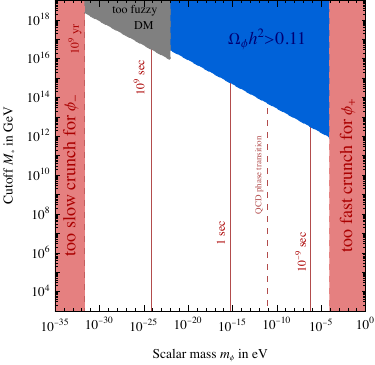} 
\caption{Parameter space of the mechanism, assuming the same mass for both scalars. The red lines denote the maximal crunching time of patches with the ``wrong" value of the weak scale. Red-shaded regions are excluded either because the crunching time is cosmologically long or because crunching would occur before the Electroweak phase transition. In the blue region oscillations of the scalars produce too much dark matter, at its boundary the cosmological DM abundance is reproduced. In the gray region the DM mass is larger than astrophysical lower bounds~\cite{Hui:2016ltb}. 
\label{fig:parameter_space}}
\end{figure}

In Fig.~\ref{fig:parameter_space} we show the allowed parameter space for $m_{\phi_+}=m_{\phi_-}$ and $\mathcal{O}_T=H_1 H_2$, which we discuss in more detail in Section~\ref{sec:pheno}. The Figure shows that cutoffs as large as $\sim M_{\rm Pl}$ can be explained by the mechanism. For cutoffs of $\mathcal{O}(M_{\rm GUT})$ coherent oscillations of the new scalars can be the DM of our Universe. The crunching time of Universes without the shallow minimum is approximately $1/m_{\phi_\pm}$. This gives an upper bound on the $\phi_+$ mass: $m_{\phi_+} \lesssim H(v) \simeq 10^{-4}$~eV. For heavier $\phi_+$ also universes with the observed Higgs vev rapidly crunch, because crunching would occur before the effect of the Higgs vev in our universe is felt by $\phi_+$. 

Note also that the lifetime of our ``safe'' metastable minimum is much longer than the age of our universe. The tunneling rate is $\Gamma/V \lesssim M_*^4 e^{-8\pi^2 M_*^2/m_{\phi_\pm}^2}$. If we take for instance $M_*\simeq 10^{14}$~GeV and $m_{\phi_+}=m_{\phi_-}=10^{-11}$~eV, a point in our Fig.~\ref{fig:parameter_space} where we also reproduce the observed DM relic density, we obtain a tunneling action $S> 8\pi^2 M_*^2/m_{\phi_\pm}^2 \sim 10^{69}$. Lowering $M_*$ all the way to a TeV and raising $m_\phi$ to $H(\Lambda_{\rm QCD})$ does not change the conclusion that our minimum is orders of magnitude more long-lived than the current age of the Universe.

\subsection{Scalar Potential and Selection of the Weak Scale}
To make the previous discussion more explicit, we consider the scalar potential  
\be
V_{\phi_-}=m_{\phi_-}^2 M^2_* \left(\frac{\phi_-}{M_*} +\frac{\phi^2_-}{2 M^2_*}-\frac{\phi^3_-}{3 M^3_*} + \frac{\delta}{4} \frac{\phi^4_-}{M^4_*}\right)+...\label{eq:V1}
\ee
and imagine that the quartic coupling is small ($\delta\ll1$). We have set to one possible numerical coefficients of the $\phi_-$ monomials, but our discussion applies also to more general choices. $V_-$ has a low-energy minimum at $\phi_- \sim M_*$, where $|V_-| \sim m_\phi^2 M^2_* \ll M^2 M^2_*$ and a deep stable minimum at $\phi_- \sim M_*/\delta$ where $-V_-\sim m_\phi^2 M^2_*/\delta^3\gtrsim M^2 M^2_*$. The potential is shown in the left panel of Fig.~\ref{fig:potential}.

This potential can naturally arise from simple supersymmetric models. We can consider for instance the superpotential
\be
W_{\phi_-} = L \Phi_- + \mu \Phi_-^2 + \lambda \Phi_-^3\, , 
\ee
and the SUSY breaking term
\be
V_B = \epsilon \mu \phi_-^3\, .
\ee
In absence of SUSY breaking, the potential from $W_{\phi_-}$ can have two widely separated minima in field space. One is at $\phi_- \sim L/\mu$ the other at $\phi_- \sim \mu/\lambda$, both have zero vacuum energy. The SUSY breaking term can split the two minima by a large amount without making the construction unnatural. In particular, for $L=m_{\phi_-} M_*$, $\mu=m_{\phi_-}$, $\epsilon=m_\phi/M_*$ and $\lambda = \sqrt{\delta} \epsilon \ll \epsilon$  we recover $V_{\phi_-}$ shifted by an unimportant overall CC of $\mathcal{O}(m_{\phi_-}^2M_*^2)$. 

This supersymmetric UV completion shows that more general choices than Eq.~\eqref{eq:V1} are natural and lead to the structure with a deep and a shallow minimum that we are interested in. In particular we do not need to consider the form in Eq.~\eqref{eq:V1} that is suggestive of the potential for a pseudo-Goldstone boson. We could take mass, cubic and tadpole at different scales. We could also consider, as we did in~\cite{TitoDAgnolo:2021nhd}, a $\mathbb{Z}_2$-symmetric potential, protected by approximate scale invariance, where the deep minimum comes from a negative quartic coupling, eventually stabilized by non-renormalizable operators at large field values. For simplicity we use Eq.~\eqref{eq:V1} in the rest of the paper, which is manifestly natural if $(m_{\phi_-}/M_*)^2 \lesssim \delta$. The second scalar that we introduced, $\phi_+$, can have the same potential as $\phi_-$, but a different sign for the cubic term
\be
V_{\phi_+}=m_{\phi_+}^2 M^2_* \left(\frac{\phi_+}{M_*} +\frac{\phi^2_+}{2 M^2_*}+\frac{\phi^3_+}{3 M^3_*} + \frac{\delta}{4} \frac{\phi^4_+}{M^4_*}\right)+... \, .
\ee
In this case the metastable minimum is not present. We only have the deep minimum at $\phi_+ \sim M_*/\delta$. The potential is shown in the left panel of Fig.~\ref{fig:potential_phip}. Clearly other possibilities are viable, but to simplify the discussion we consider the same structure for the potentials of $\phi_\pm$. In principle the vev $M_*$ and the parameter $\delta$ can be different for the two scalars, as we discussed in~\cite{TitoDAgnolo:2021nhd}. When appropriate we will comment on the impact of this possibility on phenomenology.  The last aspect that we need to specify is the coupling of $\phi_\pm$ to the SM. In this paper we will mainly consider
\be
V_{\phi_+ H}+V_{\phi_- H}= - \kappa H_1 H_2 (m_{\phi_+} \phi_+ + m_{\phi_-} \phi_-) +{\rm h.c.}\, , \label{eq:H1H2basic}
\ee
where $H_1$ is a new Higgs doublet present in addition to the SM-like Higgs $H_2$ and $\kappa\leq 1$. For $H_1 H_2$ to be a good ``trigger", i.e. select the weak scale, we need to impose an approximate $\mathbb{Z}_2$ on the Two Higgs Doublet Model (2HDM) potential. We discuss this in Section~\ref{sec:H1H2trigger}. Finally, we could consider cross-couplings between $\phi_+$ and $\phi_-$. For $\kappa \ll 1$ it is technically natural to take them to be negligibly small. Therefore, for simplicity in this paper we set them to zero, although we expect that our mechanism is effective also in the presence of cross-couplings, provided that the potential has the structure with two minima that realizes our crunching mechanism.

The mechanism can be realized also for a trigger operator $\mathcal{O}_T$ purely within the SM, as we did in~\cite{TitoDAgnolo:2021nhd}. In the following we discuss 
\be
V_{\phi_+ G}+V_{\phi_- G}=-\frac{\alpha_s}{8\pi}\left(\frac{\phi_+}{F_+} + \frac{\phi_-}{F_-}\right){\rm Tr}[G\widetilde G]\, , \label{eq:Gbasic}
\ee
expanding on the results in~\cite{TitoDAgnolo:2021nhd}. We discuss the coupling to ${\rm Tr}[G\widetilde G]$ in Section~\ref{sec:GGtrigger}, while in the following we consider the potential\footnote{For those more used to a relaxion-like parametrization of the potential: $g M^2\phi+g^2 \phi^2+...$, we note that $g=m_\phi, M_*=M^2/m_\phi$.}: 
\be
V&=&V_{\phi_+}+V_{\phi_-}+V_{H\phi_+}+V_{H\phi_-} =   m_{\phi_+}^2 M_* \phi_+ +m_{\phi_-}^2 M_*\phi_- +  \frac{m_{\phi_+}^2}{2} \phi_+^2+\frac{m_{\phi_-}^2}{2}\phi_-^2 \nn \\   
&+& \frac{m_{\phi_+}^2}{3 M_*}\phi_+^3- \frac{m_{\phi_-}^2}{3 M_*}\phi_-^3 + \delta  \frac{m_{\phi_+}^2}{4 M_*^2} \phi_+^4+ \delta \frac{m_{\phi_-}^2}{4 M_*^2} \phi_-^4  -\kappa \left( m_{\phi_+} \phi_++ m_{\phi_-} \phi_-)(H_1 H_2 +{\rm h.c.}\right)\, , \nn \\ \label{eq:Vpm}
\ee
where we recall that $\delta \ll 1$. 
The potential is technically natural for $\kappa \lesssim 4 \pi, \delta \gtrsim \max[\kappa^6 (v^4/m_{H,{\rm min}}^4), m_\phi^2/M_*^2]$ where $m_{H, {\rm min}}$ is the smallest Higgs mass in the landscape. If $m_{H, {\rm min}}\simeq 0$ the IR divergence is cutoff by $m_{\phi_\pm}$. Notice that as long as these conditions are verified, large mixed couplings are not generated by loops, at least if the parameters of the two scalars are not too different. Furthermore we will see that the values of $\kappa$ that give the observed dark matter relic density in the form of coherent oscillations of $\phi_\pm$ are $\kappa \lesssim 10^{-5}$, making induced cross couplings completely negligible. Therefore, for simplicity we can set the mixed couplings to zero, as mentioned above, to keep the analytic treatment tractable. Notice however that $\mathcal{O}(1)$ cross couplings do not necessarily spoil our mechanism, provided that at large field values they do not lift the deep minimum of $V$.

The global minimum of $V$ is at $\phi_\pm \sim \mp M_*/\delta$, where the potential is $V \sim - (m_{\phi_+}^2 +m_{\phi_-}^2)M_*^2/\delta^3$. Since the universes where the scalars are at this minimum must crunch, this is also the value of the maximal CC allowed in the landscape for our mechanism to work. For $\delta \lesssim ((m_{\phi_+}^2 +m_{\phi_-}^2)/M_*^2)^{1/3}$ this is $ \sim M_*^4$ or larger, i.e.~at the cutoff of the EFT.

The potential in Eq.~\eqref{eq:Vpm} has one metastable local minimum (where neither $\phi_+$ nor $\phi_-$ are at their global minimum) only for 
\be
\mu_S^2 \lesssim \langle H_1 H_2 \rangle \lesssim \mu_B^2\, , \label{eq:H1H1b1}
\ee
where
\be
\mu_S^2 \simeq \frac{m_{\phi_+} M_*}{\kappa}\; , \quad \mu_B^2 \simeq \frac{m_{\phi_-} M_*}{\kappa}\; . \label{eq:weak_scale}
\ee
This result can be more easily understood by considering independently the potentials for the two scalars. $V_{\phi_-}$ is depicted in the left panel of Fig.~\ref{fig:potential} and it has two cosmologically long-lived minima. If $\phi_-$ rolls to the deepest minimum the universe rapidly crunches. The coupling to the Higgs $V_{H\phi_-}$ induces a tadpole that destroys the metastable minimum at $\phi_-\sim M_*$ if $\langle H_1 H_2 \rangle \gtrsim \mu_B^2$ (right panel of Fig~\ref{fig:potential}), giving the second equality in~\eqref{eq:weak_scale}. $V_{\phi_+}$ is depicted in the left panel of Fig.~\ref{fig:potential_phip} and it has one cosmologically long-lived minimum. If $\phi_+$ rolls to the minimum the universe rapidly crunches. The coupling to the Higgs $V_{H\phi_+}$ induces a tadpole that generates a metastable minimum at $\phi_+\sim M_*$ only if $\langle H_1 H_2 \rangle \gtrsim \mu_S^2$ (right panel of Fig~\ref{fig:potential}). This gives the first equality in~\eqref{eq:weak_scale}. Only universes where this metastable minimum exists both for $\phi_\pm$ can live for cosmologically long times. These are universe where $\mu_S^2 < \langle H_1 H_2 \rangle < \mu_B^2$. 

Given our choice of trigger operator we are really selecting the vev of $H_1 H_2$. This is sufficient to select the weak scale (i.e. the vev of the SM-like Higgs) under the conditions described in Section~\ref{sec:H1H2trigger}. As shown in that Section, if we want to select the weak scale we need parametrically $\langle H_1 H_2 \rangle \simeq v^2$ which implies
\be
m_{\phi_\pm} \simeq \frac{\kappa v^2}{M_*}\, . \label{eq:mphi}
\ee
At the local minimum, if it exists, the potential is thus of order $V \simeq \kappa^2 v^4$ and the $\phi$-only potential $V_{\phi_+}+V_{\phi_-}$ is comparable to the Higgs-induced potential $V_{H\phi_+}+V_{H\phi_-}$. We imagine that the CC problem at the local minimum is solved by tuning in the landscape plus Weinberg's argument. 

\section{Cosmology}\label{sec:pheno}
In this Section we describe the cosmology of the model. The initial reheating temperature does not affect our main results. For concreteness, we take all universes to be reheated at $T\simeq M_*$. We imagine that the scalars can be in any position on their potential after reheating. In Section~\ref{sec:DM} we show that $\phi_\pm$ are good DM candidates. In Section~\ref{sec:crunching} we show that the crunching time for universes with the ``wrong" Higgs vev is dominated by the local part of the potential ($|\phi_\pm| \lesssim M_*$) and is at most $t_c \sim \max[1/m_{\phi_+}, 1/m_{\phi_-}]$.

\subsection{Dark Matter}\label{sec:DM}
As in the previous Section, we focus on the coupling to $H_1 H_2$ in Eq.~\eqref{eq:H1H2basic}. Similar results for the coupling to gluons are discussed in~\cite{TitoDAgnolo:2021nhd} and Section~\ref{sec:GGtrigger}.

The scalars $\phi_\pm$ are stable over cosmological timescales\footnote{Here $\lambda$ is an $\mathcal{O}(1)$ combination of quartics in the 2HDM Higgs sector.} 
\be
\Gamma_\phi \simeq \Gamma(\phi\to \gamma\gamma) \simeq \frac{G_F \alpha^2 m_\phi^5}{9 \sqrt{2} \pi^3 m_h^2} \left(\frac{\kappa}{\lambda}\right)^2 \simeq \frac{1}{10^{17}\times(13\times 10^9\;{\rm years})}\left(\frac{m_\phi}{\rm eV}\right)^5 \left(\frac{\kappa}{\lambda}\right)^2\, ,
\ee
and their coherent oscillations can constitute the DM of the Universe. To compute the relic density, we note that $\phi_\pm$ get a ``kick" at the Electroweak (EW) phase transition, when $T\simeq v$, and acquire an energy density in the form of a misalignment from their minimum. To be more explicit let us consider a single scalar with potential 
\be
V=V_{\phi}+V_{H\phi} &=&   m_{\phi}^2 M_*^2 \left( \frac{\phi}{M_*}+\frac{\phi^2}{2M_*^2}-\frac{\phi^3}{3M_*^3}+\delta \frac{\phi^4}{4M_*^4}\right) -\left(\kappa m_\phi \phi H_1 H_2 +{\rm h.c.}\right)\, .  \label{eq:Vphi}
\ee
Before the EW phase transition, under the conditions discussed in Section~\ref{sec:H1H2trigger} that are necessary to select the weak scale, $\langle H_1 H_2 \rangle=0$, so at early times we can focus on $V_{\phi}$. Our universe survived for cosmologically long times, so initially $|\phi|\lesssim M_*$. Universes with different initial conditions eventually see $\phi$ roll to its deep minimum and crunch independently of the value of $\langle h \rangle$. Early on, as long as $m_{\phi} \lesssim H(T)$, $\phi$ is stuck with an initial misalignment from the minimum $\phi_I$ and an energy density given by $V_\phi \simeq m_\phi^2 \phi_I^2 \lesssim m_{\phi}^2 M_*^2$. When $m_{\phi} \gtrsim H(T)$ it starts to oscillate around its metastable minimum $\phi_{\rm min} \simeq M_*$, and its energy density starts to redshift like cold DM. This can occur either before ($m_{\phi} \gtrsim H(v) \simeq 10^{-5}$~eV) or after ($m_{\phi} \lesssim H(v)$) the EW phase transition. We can call $\phi_{\rm EW}$ the average amplitude of the field at the EW phase transition. This is given by $\phi_{\rm EW}=\phi_I$ if $m_{\phi} \lesssim H(v)$ and $\phi_{\rm EW}=\phi_I (a(T_{\rm osc})/a(v))^{3/2}$ if $m_{\phi} \gtrsim H(v)$, where $a(T)$ is the scale factor of our universe and $T_{\rm osc}$ the temperature at which $\phi$ starts to oscillate. In both cases $|\phi_{\rm EW}|\lesssim M_*$.

At the EW phase transition the average position of $\phi$ in its potential is $\bar \phi \simeq M_*+\phi_{\rm EW} \simeq M_*$  and $V_{H\phi}$ starts contributing to the $\phi$ potential 
\be
\Delta V=V_{H\phi}\simeq \kappa m_\phi M_* \langle H_1 H_2 \rangle_{\rm us}\simeq \kappa m_\phi M_* v^2 \label{eq:DeltaV}\, ,
\ee
where $\langle H_1 H_2 \rangle_{\rm us}$ is the operator vev in our universe. If our universe is close to one of the boundaries of the ``safe" region for the Higgs vev, i.e. $\langle H_1 H_2\rangle_{\rm us} \simeq \mu_{S}^2$ or $\langle H_1 H_2\rangle_{\rm us}\simeq \mu_B^2$, then 
$\Delta V \simeq V_\phi(M_*)$ and the minimum of $\phi$ is shifted from its initial position, 
\be
\Delta \phi_{\rm min} \simeq M_*\, . \label{eq:Deltaphi}
\ee
This contributes another factor of $M_*$ to $\phi$'s initial misalignment. Generically we expect to be in the situation $\langle H_1 H_2\rangle_{\rm us}\simeq \mu_B^2$, given the distribution of mass squared parameters in a typical landscape (i.e. since we need to tune to make $\langle H_1 H_2\rangle$ small, larger values are generically preferred). Therefore in the following we imagine that $\langle H_1 H_2\rangle_{\rm us}\simeq \mu_B^2$ when $\langle h \rangle \simeq v$ and take Eq.~\eqref{eq:Deltaphi} as a good parametric estimate of the misalignment of $\phi_-$ at the EW phase transition. If $M_*$ and $\kappa$ are the same for both scalars, $\phi_+$ gives at most a comparable contribution to the DM relic density, and only if it starts oscillating and redshifting as cold DM after the EW phase transition. Since we are interested in a first estimate of the relic density, we neglect the $\phi_+$ relic density and continue with our single scalar description, which captures the relevant parametrics.

\begin{figure}[t]
\includegraphics[width=0.6\textwidth]{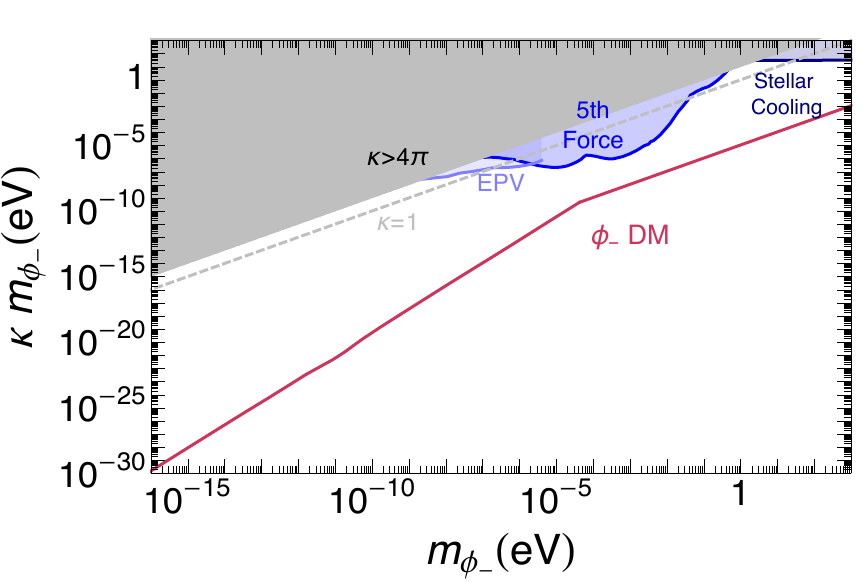} 
\caption{Laboratory and astrophysical constraints on a scalar coupled to the Higgs boson via the trilinear interaction $\kappa m_{\phi_-} \phi_- |H|^2$. The bounds include tests of the equivalence principle~\cite{Smith:1999cr, Schlamminger:2007ht,Berge:2017ovy}, tests of the Newtonian and Casimir potentials (5th force)~\cite{Spero:1980zz, Hoskins:1985tn, Chiaverini:2002cb, Hoyle:2004cw, Smullin:2005iv, Kapner:2006si,Bordag:2001qi, Bordag:2009zzd, Turyshev:2006gm} and stellar cooling constraints~\cite{Hardy:2016kme}. The red solid line shows the target given by $\phi_-$ reproducing the observed dark matter relic density. Above the gray dashed line $\kappa > 1$. In the gray shaded region $\kappa > 4\pi$, making the scalar potential unnatural. The constraint from AURIGA~\cite{Branca:2016rez} is not shown because the mass ranged explored is too narrow to be visible on this scale. The bound does not touch our DM parameter space. Bounds on this coupling and future probes, spanning a larger mass range, can be found in~\cite{Flacke:2016szy, Banerjee:2020kww}.}
\label{fig:DM}
\end{figure}

The kick at the EW phase transition gives the dominant contribution to the relic density if $m_\phi \gtrsim H(v)$, since the initial misalignment (that can be at most $\mathcal{O}(M_*)$) has already partially redshifted away. If $m_\phi \lesssim H(v)$, ignoring the initial misalignment still gives parametrically the correct result, since it can give at most an $\mathcal{O}(1)$ correction on top of the EW-induced misalignment. Therefore modulo $\mathcal{O}(1)$ factors, we get
\be
\rho_{\phi_-}(T\simeq v)\simeq m_\phi^2 M_*^2 \gtrsim \rho_{\phi_+}(T\simeq v)\, . 
\ee
From Eq.~\eqref{eq:mphi} we know that to select the weak scale we need $m_\phi^2 M_*^2 \simeq \kappa^2 v^4$, so the relic density is entirely specified by giving the coupling $\kappa$ of the scalars to the SM, and their mass $m_{\phi}$, which determines the moment in time when they start to oscillate and redshift as cold DM ($m_\phi \simeq H(T_{\rm osc})$). We are in the same situation described in~\cite{Arkani-Hamed:2020yna, TitoDAgnolo:2021nhd}. Light scalars coupled to trigger operators offer universal targets to DM searches. We now give an estimate of the target. The relic density today is
\be
\frac{\rho_{\phi_-}+\rho_{\phi_+}}{\rho_{\rm DM}}\simeq \frac{\rho_{\phi_-}}{\rho_{\rm DM}} \simeq  m_{\phi_-}^2 M^2_* \frac{s_0}{\rho_{\rm DM}^0 \min[s(v), s(T_{\rm osc})]}\, .
\ee
To highlight the phenomenological significance of this result we can use Eq.~\eqref{eq:mphi}: $m_{\phi_-}^2 M^2_*\simeq \kappa^2 v^4$ and rewrite our expression in terms of the effective trilinear coupling of $\phi_-$ with the Higgs that determines the strength of $\phi_-$ interactions with the SM:
\be
\mathcal{L}\supset - b_- \phi_- H_1 H_2 +{\rm h.c.} \simeq - b_- \phi_- |H|^2 +..., \quad b_- \simeq \kappa m_{\phi_-}\, .
\ee
Here for simplicity we have taken the limit of a small coupling of $H_1$ to SM fermions (i.e. $\lambda_3+\lambda_4 +\lambda_5 \ll \lambda_2$ with $\lambda_i$'s defined in Eq.~\eqref{eq:H1H2}; generalizing introduces additional $\mathcal{O}(1)$ factors that do not qualitatively affect our discussion). In conclusion
\be
\frac{\rho_\phi}{\rho_{\rm DM}}= \frac{b_-^2 v^4}{m_{\phi_-}^2} \frac{s_0}{\rho_{\rm DM}^0 \min[s(v), s(T_{\rm osc})]} \simeq \left\{\begin{array}{c}\frac{b^2 v}{m_{\phi_-}^2 T_{\rm eq}}\quad m_{\phi_-} \geq H(v) \\ \frac{b^2 v^4}{m_{\phi_-}^{7/2} M_{\rm Pl}^{3/2} T_{\rm eq}}\quad m_{\phi_-} < H(v)\end{array}\right.
\ee
where $T_{\rm eq}\simeq {\rm eV}$ is the temperature of matter-radiation equality, and we have a target for ultralight DM searches:
\be
b_{\rm DM} \simeq m_{\phi_-} \sqrt{\frac{T_{\rm eq}}{v}}\min\left[1,\frac{m_{\phi_-}^{3/2}M_{\rm Pl}^{3/2}}{v^3}\right]\, .
\ee
For any given mass only one value of the coupling to the SM $b_{\rm DM}$ gives the observed relic density. In Figure~\ref{fig:DM} we show this ultralight DM target and current constraints on our parameter space. The bounds include tests of the equivalence principle~\cite{Smith:1999cr, Schlamminger:2007ht,Berge:2017ovy}, tests of the Newtonian and Casimir potentials (5th force)~\cite{Spero:1980zz, Hoskins:1985tn, Chiaverini:2002cb, Hoyle:2004cw, Smullin:2005iv, Kapner:2006si,Bordag:2001qi, Bordag:2009zzd, Turyshev:2006gm} and stellar cooling~\cite{Hardy:2016kme}.

Future probes of $\phi_-$ dark matter, including torsion balance experiments~\cite{Graham:2015ifn}, atom interferometry~\cite{Arvanitaki:2016fyj}, optical/optical clock comparisons  and nuclear/optical clock comparisons~\cite{Arvanitaki:2014faa}, resonant mass detectors (DUAL and SiDUAL~\cite{Leaci:2008zza}) and gravitational-wave detectors~\cite{Grote:2019uvn,Vermeulen:2021epa} are orders of magnitude too weak to probe our parameter space. Current constraints on 5th forces that are more than twenty years old are relatively close to motivated parameter space in the range $10^{-5}\;{\rm eV} \lesssim m_{\phi_-} \lesssim 10^{-3}\;{\rm eV}$ and we hope that this study will motivate future efforts towards improving their sensitivity. 

Modulo factors related to the multiplicity of scalars, the prediction for the relic density is exactly the same as in~\cite{Arkani-Hamed:2020yna} and similar considerations can be made in relaxion models~\cite{Banerjee:2018xmn}. This is one manifestation of the universality of this prediction. Light scalars that can select the weak scale, generically get the biggest contribution to their relic density from a SM phase transition. If the Universe is reheated above the relevant phase transition, their relic density today depends only on their mass and coupling to the SM.

\subsection{Crunching dynamics}\label{sec:crunching}

\begin{figure}[t]
\includegraphics[width=0.45\textwidth]{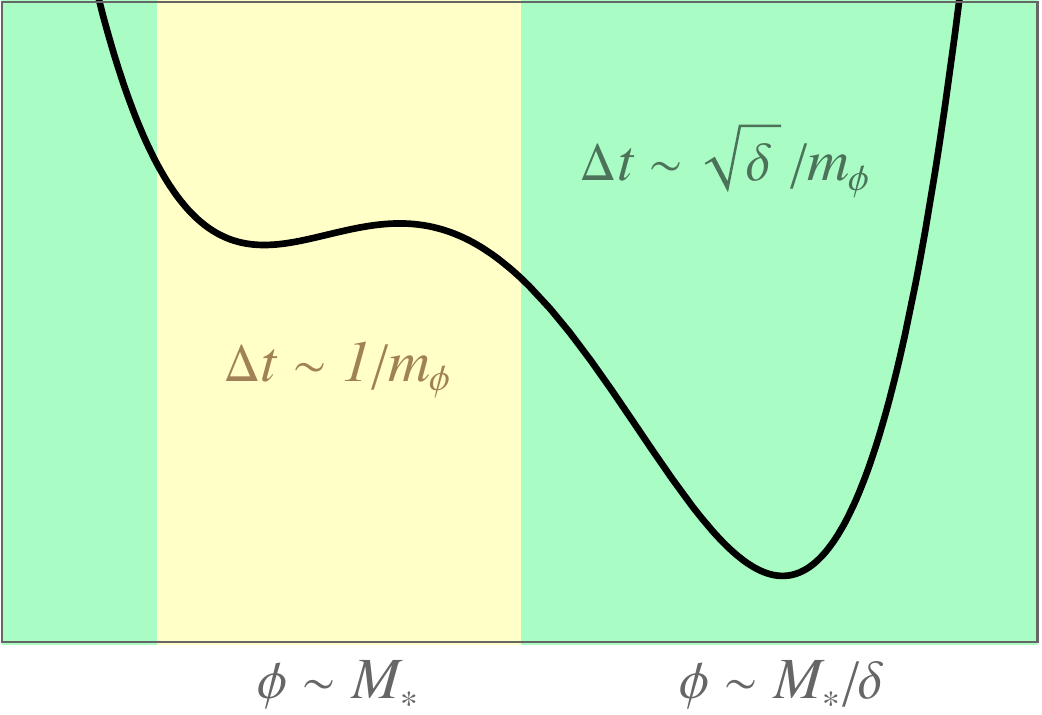} 
\caption{Schematic view of the structure of the potential including the time to cross the local (global) region around (far from) the metastable minimum. Scalars that roll to the deep minimum and lead to a crunching universe take most of the time to cross the local region ($\sqrt{\delta}\ll 1$).}
\label{fig:crunching}
\end{figure}

In this Section we consider the dynamics of $\phi_\pm$ crunching in detail and calculate the crunching time. We follow the evolution of the Universe after inflation, starting from a SM reheating temperature of the order of the cutoff, $T \sim M_*$. If the reheating temperature is lower than this, similar considerations are possible. 

We want to solve the classical equations of motion in an expanding universe
\be
\ddot \phi + 3 H \dot \phi + \frac{\partial V}{\partial \phi}=0\, , \label{eq:EOMs}
\ee
for both $\phi_+$ and $\phi_-$, assuming that initially $\dot \phi_\pm(t_0)=0$, $T\simeq M_*$. Since the two scalars are approximately decoupled ($\kappa \lesssim 10^{-5}$ to get the observed DM relic density) we can solve Eq.~\eqref{eq:EOMs} separately for $\phi_+$ and $\phi_-$. As in the DM case we can consider a single scalar $\phi$, solve its equations of motion and then see how the solution applies to $\phi_+$ and $\phi_-$.

In principle there are four relevant regimes (that if needed can be glued together). They correspond to the position of $\phi$ (near the local minimum or as far as it can be, see Fig.~\ref{fig:crunching}) and to whether $H(T)$ is dominated by the $\phi_\pm$ vacuum energy or SM radiation. If $\phi_\pm$ vacuum energy dominates the expansion of the universe the patch is in a state of $\phi_\pm$-driven inflation until the rolling of the scalars makes it crunch. These patches do not reheat\footnote{More precisely, as described in~\cite{Strumia:2019kxg}, while the scalar slides down its potential a subdominant thermal bath is formed, due to the tiny interaction with the SM photons. When the vacuum energy crosses zero and crunching starts, both the kinetic energy of $\phi$ and the thermal bath rapidly blue-shift until the big crunch.}, because of the feeble $\phi_\pm$ interactions. As a consequence, independently on the rolling time, these patches are basically empty, excluded by the standard anthropic arguments on the possibility of complex structures. In summary only two cases are actually relevant:
\begin{enumerate}
\item $H(T)\simeq T^2/M_{\rm Pl}$, $\phi(t_0) \simeq M_*$ 
\item $H(T)\simeq T^2/M_{\rm Pl}$, $\phi(t_0) \simeq M_*/\delta$ . 
\end{enumerate}
First consider patches that start with the fields $\phi_{\pm}$, denoted generically by $\phi$, at the scale of the local minimum of the potential, i.e. $\phi \sim M_{*}$. Since at this scale $V \sim m_\phi^2 M_*^2 \ll M^2 M_*^2$, the patch is initially radiation-dominated and the evolution of the scalars is given by
\be \label{eq:EoMlocal}
\ddot \phi + \frac{3}{2 \, t} \dot\phi + \frac{\partial V}{\partial \phi} = 0 \;. \label{eq:EOM}
\ee
In universes destined to crunch and in the local region $|\phi| \lesssim M_*$, we can approximate $V$ with a tadpole (either the Higgs induced one for $\phi_-$ or the one in $V_{\phi_+}$ for $\phi_+$), so $\frac{\partial V}{\partial \phi} \simeq {\rm const.}$ and we can solve Eq.~\eqref{eq:EOM} exactly. We find that $\phi_\pm$ cross a region of $\mathcal{O}(M_*)$ in a time 
\be
\Delta t_-(M_*) &=& \frac{\sqrt{5}}{2\sqrt{\kappa (m_{\phi_-}/M_*)}\mu_H}\, , \quad \mu_H^2 \equiv \langle H_1 H_2 \rangle\, , \nn \\
\Delta t_+(M_*) &=& \frac{\sqrt{5}}{2m_{\phi_+}}\, ,
\ee
respectively. The longest crossing time for universes with $\langle h\rangle \gtrsim v$ is obtained for $\langle H_1 H_2 \rangle\simeq \mu_B^2$, i.e. when the Higgs-induced tadpole has the smallest slope that can still destroy the local minimum for $\phi_-$. This happens for $\Delta t_-(M_*) \simeq 1/m_{\phi_-}$.

 To make sure that this calculation is consistent we need to check that in a time $\Delta t_\pm$ the temperature has not dropped enough from the initial value to take the universe to a new phase of inflation. We have
\be
\Delta t_\pm = - \int_{M_*}^{T_\pm} \frac{dT}{H(T) T} \rightarrow T_\pm^4 = \frac{M^4_*}{(2H(M_*) \Delta t_\pm+1)^2}\, .
\ee
Comparing with $V_{\phi_\pm} \simeq m_{\phi_{\pm}} M_*^2$ we get that we do not enter a phase of inflation if 
\be
\frac{H(M_*)}{M_*} \lesssim 1\, ,
\ee
which is satisfied for sub-Planckian $M_*$. Now, let us instead assume that the patch starts from a value of the scalar fields at the global scale, i.e. $\phi \sim M_*/\delta$ and $V \sim m_\phi^2 M_*^4/\delta^3$. If $\delta^3 \lesssim m_\phi^2/M_*^2$ the scalar potential dominates with respect to the thermal bath and the patch is in a state of $\phi$-driven inflation until the rolling of the scalars makes it crunch. As explained above these universes are empty. 

In the opposite regime $\delta^3 \gtrsim m_\phi^2/M_*^2$ the patch starts as radiation-dominated. After a time $\Delta t_s \lesssim \sqrt{\delta}/m_\phi$ Hubble friction becomes negligible\footnote{Hubble friction can be negligible from the beginning for low cutoffs, i.e. if $H(M_*) \simeq M_*^2/M_{\rm Pl} \lesssim m_\phi/\sqrt{\delta}.$}. This can be estimated for instance by showing that when $H(T) \simeq m_\phi/\sqrt{\delta}$, $\phi$ is slow rolling over a range $\Delta \phi \simeq M_*/\delta$ in one Hubble time. When Hubble friction is negligible we can solve Eq.~\eqref{eq:EOMs} in its simpler form
\be
\ddot \phi + \frac{\partial V}{\partial \phi}\simeq 0\, .
\ee
We obtain that $\phi$ crosses the ``global" region $\Delta \phi \simeq M_*/\delta$, in a time $\Delta t_g \simeq \sqrt{\delta}/m_\phi$. Therefore the longest time that $\phi$ can spend in this region of the potential, obtained combining the two times (the time in which $\phi$ can be stuck due to Hubble friction and the time needed to cross the region), $\Delta t_g+\Delta t_s\simeq \sqrt{\delta}/m_\phi$, is much shorter than the one required to cross the region around the local minimum: $\Delta t_\pm \lesssim 1/m_\phi$. 

In summary, as shown in Fig.~\ref{fig:crunching}, the longest time that it can take a universe with the wrong Higgs vev to crunch is parametrically
\be
\Delta t_c^{\rm max} \simeq \max[1/m_{\phi_+}, 1/m_{\phi_-}]\, , \label{eq:tc}
\ee
dominated by patches where $\phi_\pm$ are initially in the region where their local minimum can be generated $|\phi_\pm|\lesssim M_*$.

Finally, notice that: 1) If $\Delta t_g$ is consistently smaller than a Hubble time, the global region is crossed in a time $\Delta t_g$ and the crunching time is dominated by the time to cross the local region, as discussed above. In the opposite case, instead, the temperature drops until $T^4 \sim  M_{\rm Pl}^2m_\phi^2/\delta$. This can be smaller than $V$ itself, signalling the onset of a stage of $\phi$-driven inflation, which would give an empty patch till crunching. 2) a patch starting from sufficiently far away from the local minimum could be doomed to crunch anyway, independently on the value of the Higgs vev, since the kinetic energy of $\phi$ when Hubble friction becomes negligible, which for $\phi$ initially at $M_*/\delta$ is given by 
\be
\dot \phi^2 \simeq \frac{m_\phi^2 M_*^2}{\delta^3}\, ,
\ee
can be sufficient to overtake the local maximum and access the unstable region of the potential. Nevertheless, the crunching time of these patches is at most the one given in Eq.~\eqref{eq:tc}, so our mechanism is effective as long as $t_c \sim 1/m_\phi$ is short enough. 

In conclusion some patches might crunch or enter a phase of $\phi$-driven inflation, leading to an empty universe, even if they have the observed Higgs vev. However all patches with the wrong Higgs vev rapidly crunch or enter a phase of $\phi$-driven inflation, in a time bounded by~\eqref{eq:tc}, making our mechanism an effective way to select the weak scale. 

\subsection{Parameter space}

Our parameter space is summarized in Fig.~\ref{fig:parameter_space}. The scalar mass $m_\phi$ is bounded from below by the requirement that the crunching time must be shorter than the cosmological scale, say $10^9 \,\unit{yr}$, otherwise patches with heavy Higgs, or without EW symmetry breaking, are too long-lived. Imposing a shorter maximum crunching time a more stringent limit is obtained, as shown in Fig.~\ref{fig:parameter_space}. On the other hand, $m_\phi$ is bounded from above by the requirement that the crunching time for $\phi_+$ must be longer than $1/H$ at the EW phase transition, so that the the Higgs vev has the possibility to stop the rolling of $\phi_+$ in due time. Finally, the cutoff $M_*$ is bounded from above by the requirement that scalar oscillations do not overclose the Universe. If $M_* \gtrsim 10^{12} \,\unit{GeV}$, they can reproduce the observed DM relic density. If this is the case, the scalars must  however be heavier than $\approx 10^{-22} \, \unit{eV}$, because of limits on fuzzy DM~\cite{Hui:2016ltb}.

\section{The $H_1H_2$ trigger}\label{sec:H1H2trigger}

The essence of our mechanism is the generation of a Higgs-dependent tadpole for two scalars $\phi_{\pm}$. When the Higgs vev is larger than a certain threshold, $\langle h \rangle \gtrsim \bar \mu_S$, this tadpole generates a ``safe" minimum for $\phi_+$. When it gets even larger, $\langle h \rangle \gtrsim \bar\mu_B \gtrsim \bar\mu_S$, it destabilizes a minimum for $\phi_-$. As discussed in Section~\ref{sec:mechanism} and Section~\ref{sec:pheno}, only universes with the Higgs vev in the range $\bar\mu_S \lesssim \langle h \rangle \lesssim \bar\mu_B$, do not rapidly crunch. So far we have mainly considered one operator that can generate this Higgs-dependent tadpole
\be
\mathcal{O}_{T}=H_1 H_2\, , \quad \mathcal{L}\supset - \kappa H_1H_2 (m_{\phi_+} \phi_+ + m_{\phi_-}\phi_- ) +{\rm h.c.}\, . \label{eq:Htadpole}
\ee
This type of operator is a {\it trigger} in the definition of~\cite{Arkani-Hamed:2020yna}. When the Higgs vev (and thus the operator vev) crosses certain upper or lower bounds, a cosmological event is triggered via the coupling to the new scalar(s). In our case the event is a rapid crunch of the universe. 

In this Section we discuss the dependence of $\langle H_1 H_2\rangle$ on the vev of the SM Higgs in more detail. In particular we show how bounding the vev of $H_1 H_2$ selects a value for $\langle h \rangle$ if the Two Higgs Doublet Model (2HDM) has a $\mathbb{Z}_2$ symmetry: $H_1 H_2 \to - H_1 H_2$. This singles out a very specific kind of 2HDM potential that leads to characteristic signals at the LHC. We find interesting that discovering new fundamental scalars at the LHC, without new symmetries protecting their masses, is traditionally considered as a ``death sentence" for naturalness. On the contrary, our study and the work in~\cite{Arkani-Hamed:2020yna} show that this can be the first manifestation of naturalness of the Higgs mass.

We consider the most general $\mathbb{Z}_2$ symmetric 2HDM potential~\cite{Arkani-Hamed:2020yna}
\be
&&H_1 H_2 \to - H_1 H_2\, , \quad  V_{H_1 H_2}\to V_{H_1 H_2}\, ,
\ee
where
\be
V_{H_1 H_2}&=& \frac{m_1^2}{2} |H_1|^2 + \frac{m_2^2}{2} |H_2|^2 +\frac{\lambda_1}{2}|H_1|^4 +\frac{\lambda_2}{2}|H_2|^4 \nn \\ &+&\lambda_3 |H_1|^2|H_2|^2 +\lambda_4 |H_1 H_2|^2 +\left(\frac{\lambda_5}{2}(H_1 H_2)^2 +{\rm h.c.}\right)\, .
\label{eq:H1H2}
\ee
This potential does not contain odd spurions that can generate contributions to $\mu_H^2=\langle H_1 H_2\rangle$ sensitive to the cutoff. If the $\mathbb{Z}_2$ is exact we have $\mu_H^2=0$. Coupling $H_1 H_2$ to $\phi_\pm$, as in Eq.~\eqref{eq:Htadpole}, does not break the $\mathbb{Z}_2$ symmetry if $\phi_\pm \rightarrow - \phi_\pm$. Furthermore it leaves $\mu_H^2$ and all 2HDM phenomenology approximately unaltered, since the couplings between the new scalars and the 2HDM are minuscule $m_\phi\lesssim v^2/M_*$, as discussed in Sections~\ref{sec:mechanism} and~\ref{sec:pheno}. Therefore, in the study of $\mu_H^2$ we can ignore the coupling to $\phi_\pm$ and only~\eqref{eq:H1H2} matters.

The vevs of $H_{1,2}$ and the QCD condensate break the $\mathbb{Z}_2$ and generate $\mu_H^2\neq 0$. To compute the value of $\mu_H^2$ we need to assign $\mathbb{Z}_2$ charges to the quarks and leptons. We choose
\be
H_2 \to - H_2, \quad (qu^c)\to -(qu^c), \quad (qd^c)\to -(qd^c), \quad (le^c)\to -(le^c), 
\ee
so that one of the two Higgs doublets is inert and the only Yukawa couplings in the model are
\be
V_Y=Y_u q H_2 u^c +Y_d q H_2^\dagger d^c+Y_e l H_2^\dagger e^c +{\rm h.c.}\, . \label{eq:Y}
\ee
This is the safest choice phenomenologically. It was shown in~\cite{Arkani-Hamed:2020yna} that this charge assignment is still viable experimentally, but it will be decisively probed by HL-LHC. 

\begin{figure}[t]
\includegraphics[width=0.45\textwidth]{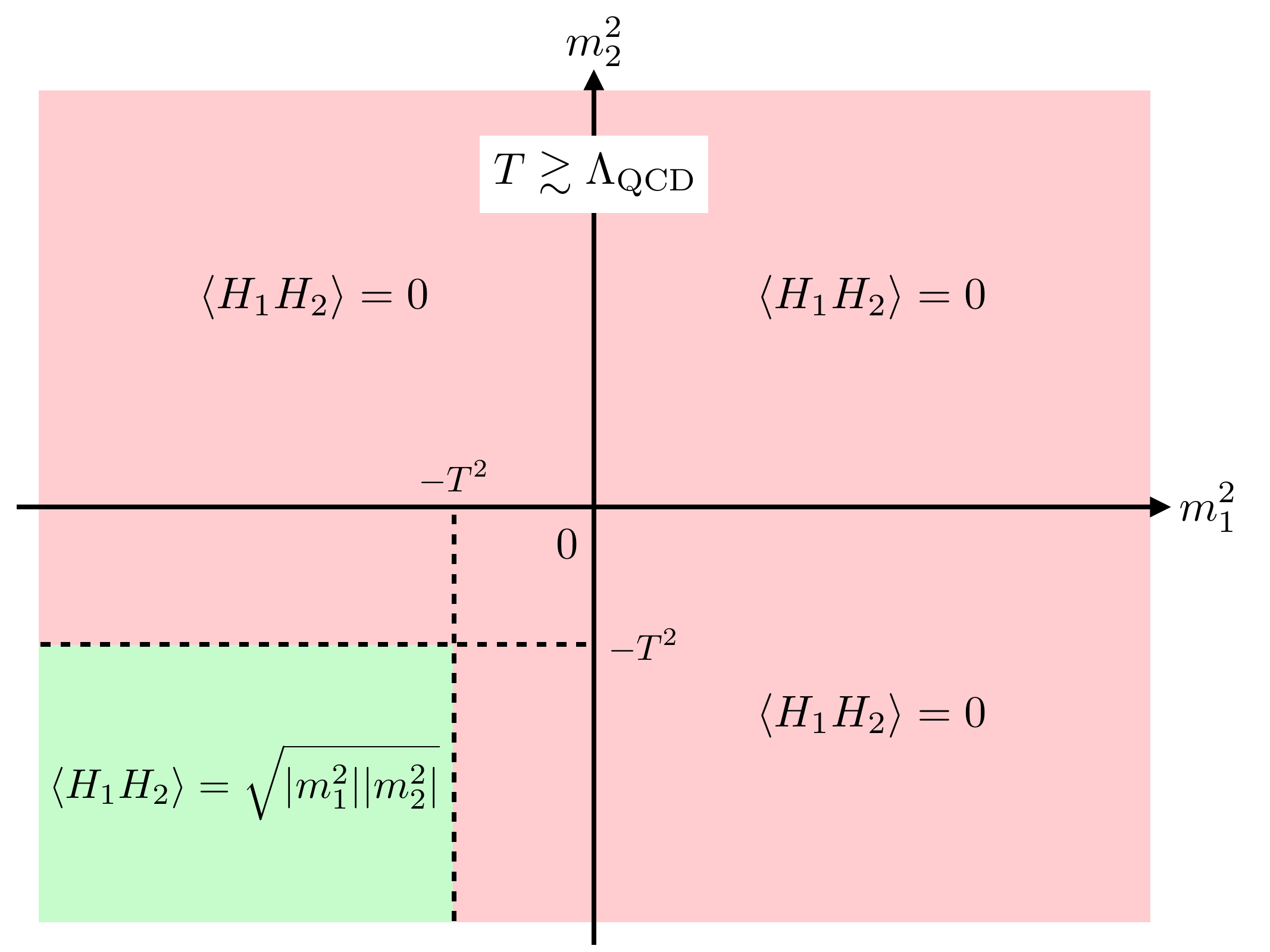} 
\includegraphics[width=0.45\textwidth]{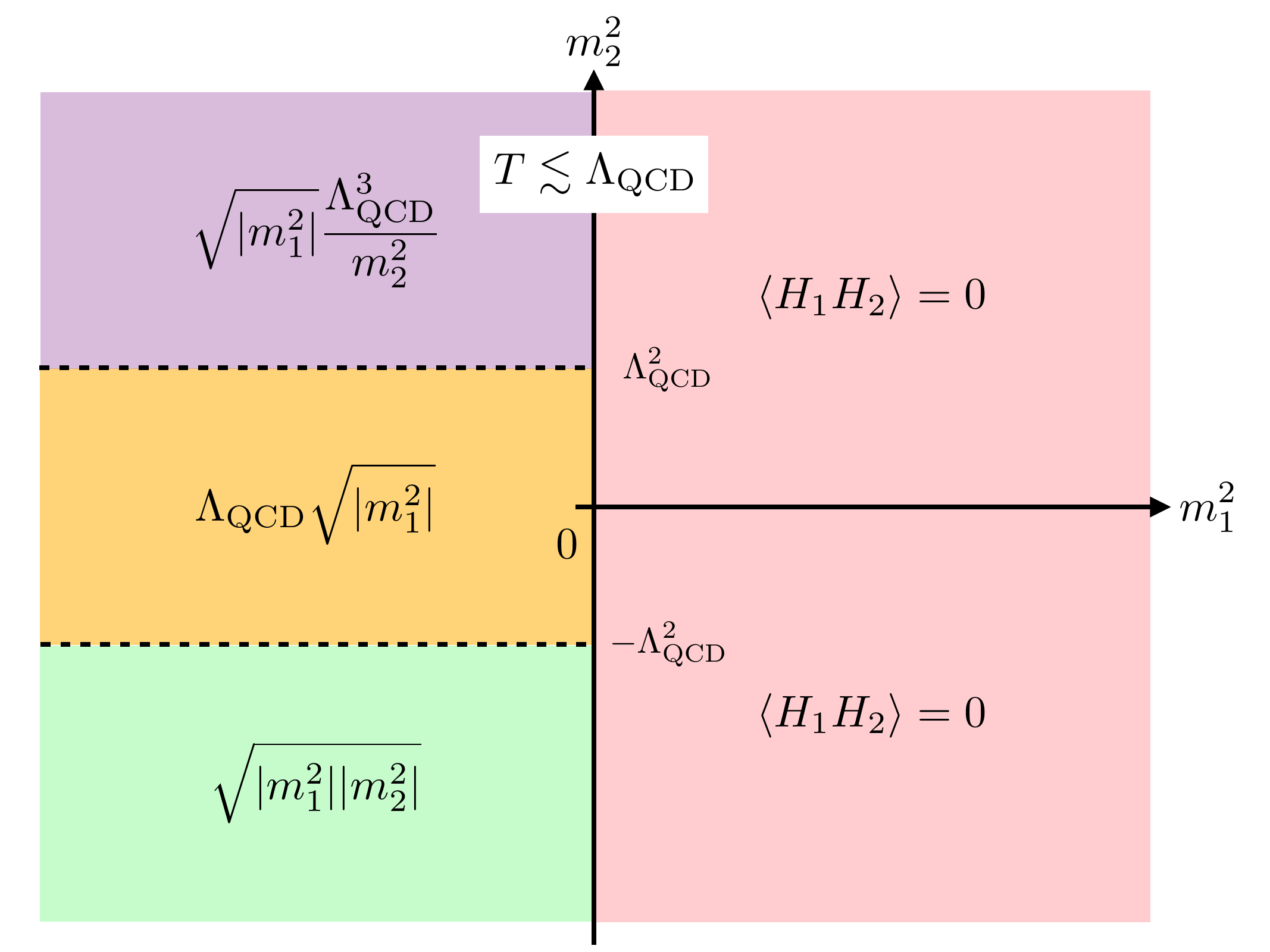} 
\caption{Vacuum expectation value of $H_1 H_2$ in the model of Eq.s~\eqref{eq:H1H2} and \eqref{eq:Y} as a function of the two Higgs masses. We show the vev before (left) and after (right) the QCD phase transition. $m_{1,2}$ are effective masses with the dimensions of vevs that contain contributions from $\mathcal{O}(1)$ quartic couplings. $\Lambda_{\rm QCD}$ is a function of the Higgs vevs and varies within the purple and yellow boxes.  We have approximated thermal corrections to $m_{1,2}^2$ with $T^2$ to improve readability. Note that in the light red regions the vev is not exactly zero, because of a small effective $B\mu$ term induced by the $\phi_\pm$ vevs. However this effect is too small to affect our conclusions. It only gets rid of dangerous domain walls, as discussed in the main body of the text.}
\label{fig:vev}
\end{figure}

The model defined by Eq.s~\eqref{eq:H1H2} and \eqref{eq:Y} has a UV-insensitive and calculable vev $\mu_H^2$, shown in Fig.~\ref{fig:vev}. $\mu_H^2$ gives a tadpole to $\phi_\pm$ and so the mechanism is really selecting
\be
\mu_S^2 \lesssim \langle H_1 H_2 \rangle \lesssim \mu_B^2\, ,
\ee
which is not the vev of the SM Higgs: $\langle h \rangle \simeq \sqrt{|m_2^2|}$. In principle $\mu_H^2$ can be close to $v^2$ also for universes with very different EW-symmetry breaking compared to ours, for instance $m_1^2 \sim -  v^4/M^2_*$, $m_2^2 \sim -M^2_*$ still gives $\mu_H^2(T=0)\simeq v^2$. However our selection mechanism takes place at $T\neq 0$. In practice we never need to worry about these patches provided that $\phi_+$ is heavy enough to roll to its stable minimum before $H_1 H_2$ gets a vev in these universes. In our universe $\mu_H^2 \neq 0$ already at the EW phase transition, while in these patches it is zero until much later: $T ^2\lesssim v^4/M^2_*$.

There is one additional subtlety to consider. QCD can generate a vev for $H_1 H_2$ even for $m_2^2\geq0$ (see the right panel of Fig.~\ref{fig:vev}): at the QCD phase transition quark bilinears condensate. This gives an effective tadpole for $H_2$, via~\eqref{eq:Y}. As a consequence, $\mu_H^2$ can be close to $v^2$ also for another class of universes with very different EW-symmetry breaking compared to ours, for instance $m_1^2 \sim -  v^4/\Lambda_{\rm QCD}^2$, $m_2^2 \simeq 0$ still gives $\mu_H^2(T=0)\simeq v^2$. For concreteness here and in the following we assume that dimensionless couplings do not scan in the landscape. $\Lambda_{\rm QCD}$ is still different from universe to universe due to the different SM Higgs vevs, but this does not affect our discussion, so we do not show explicitly this dependence here and in the following.

As in the previous case, these unwanted patches rapidly crunch if $\phi_+$ is heavy enough to roll to its stable minimum before the QCD phase transition. Indeed, as shown in the left panel of Fig.~\ref{fig:vev}, before the QCD phase transition $\mu_H^2$ can be nonzero only if both $m_{1,2}^2 < 0$ and larger, in absolute size, than the positive thermal contribution.  These considerations favour a relative heavy $\phi_+$, close to the boundary of its allowed region $m_{\phi_+} \lesssim H(v) \simeq 10^{-4} \, \textrm{eV}$.

There are other possibilities to solve the problem raised by these unwanted patches (both those with $m_2^2>0$ and those with small $m_1^2$ or $m_2^2$). We can consider low cutoffs $M_* \lesssim v^2/\Lambda_{\rm QCD}$, so that these patches are not present in the Multiverse or supplement the mechanism with the anthropic considerations in~\cite{Arkani-Hamed:2020yna}.

The model that we just described has an accidental symmetry, as noted in~\cite{Arkani-Hamed:2020yna}. The Lagrangian is actually $\mathbb{Z}_4$-symmetric
\be
H_1 \to i H_1 \quad H_2 \to i H_2 \quad (qu^c)\to -i(qu^c) \quad (qd^c)\to i(qd^c) \quad (le^c)\to i(le^c)\, , \label{eq:Z4}
\ee
this creates a potential cosmological problem. After EW symmetry breaking a $\mathbb{Z}_2$ subgroup of the $\mathbb{Z}_4$ survives
\be
H_1 \to - H_1\, .
\ee
This $\mathbb{Z}_2$ subgroup can be obtained from the $\mathbb{Z}_4$ after a global hypercharge rotation. As a consequence the model has a domain-wall problem, i.e. domain walls between regions with $\pm v_1$ are generated at the EW phase transition and they come to dominate the energy density of our Universe at $T\simeq v(v/M_{\rm Pl})^{1/2}\simeq $~keV. We can solve the problem via a tiny breaking of the $\mathbb{Z}_4$ that does not alter any of our conclusions. If the 2HDM potential contains a $B\mu$-term of size
\be
V_{H_1 H_2}\supset - B\mu H_1 H_2+{\rm h.c.} \quad B\mu \simeq \frac{v^4}{M_{\rm Pl}^2}\, . \label{eq:Bmu}
\ee
This insures that the domain walls annihilate at $T\simeq$~keV. At larger temperatures they constitute a negligibly small fraction of the total energy density~\cite{Arkani-Hamed:2020yna}. This $B\mu$ term breaks also our original $\mathbb{Z}_2$, but it is numerically negligible in our analysis. In a large fraction of our parameter space, shown in Fig.~\ref{fig:parameter_space}, the misalignment of $\phi_\pm$ at the EW phase transition automatically generates a large enough $B\mu$, and we do not need Eq.~\eqref{eq:Bmu}. 

As noted in~\cite{Arkani-Hamed:2020yna} the phenomenology of this $\mathbb{Z}_2$ symmetric ``type-0" 2HDM is very interesting. Since we effectively set to zero any scale in the potential besides the two masses ($B\mu \simeq \frac{v^4}{M_{\rm Pl}^2}\ll v^2$), the new Higgs states contained in $H_1$ are close to the weak scale. If we adopt the usual notation for charged, scalar and pseudo-scalar Higgses we have
\be
m_A^2 &=& - v^2 \lambda_5\, , \nn \\
m_{H^\pm}^2 &=& - v^2 \frac{\lambda_5+\lambda_4}{2} \nn \\
m_{h,H}^2 &=& \frac{1}{2}\left(\lambda_1 v_1^2+\lambda_2 v_2^2\pm\sqrt{\left(\lambda_2 v_2^2-\lambda_1 v_1^2\right)^2+4 v_1^2v_2^2\lambda_{345}^2}\right) \label{eq:masses}
\ee 
and to avoid TeV-scale Landau poles we need all quartics to be $\lesssim 2$ around the weak scale~\cite{Arkani-Hamed:2020yna}. Therefore we 
have a sharp target for searches at the LHC and HL-LHC, which is made even sharper if we notice two well-known facts: 1) There are couplings between the SM and two new Higgses proportional to the $SU(2)_L$ gauge coupling, which are fixed by gauge invariance. 2) Couplings with a single new Higgs, that are proportional to $v_1$, can not be made arbitrarily small. 

Both points are quite interesting for the LHC: 1) a CMS search for staus in a $\tau^+\tau^-+$MET final state, is sensitive to pair production of $H^+H^-$~\cite{CMS:2019eln}. If recasted it can potentially extend LEP's bound on the $H^\pm$ mass to about 150 GeV~\cite{Arkani-Hamed:2020yna}. 2) At small $v_1$ the new scalar Higgs becomes light
\be
m_H^2 &=& v_1^2 \left(\lambda_1 - \frac{\lambda_{345}^2}{\lambda_2}\right)+\mathcal{O}(v_1^4/v^4)\, .
\label{eq:massesII}
\ee
So when trying to decouple $H_1$ we rapidly run into stringent constraints from LEP, B-factories and beam-dump experiments. Quantitatively this means that Higgs coupling deviations in this model will be visible at HL-LHC. A more complete summary of signals and constraints can be found in~\cite{Arkani-Hamed:2020yna}.

To conclude this section it is worth to point out that the $\mathbb{Z}_2$ symmetry is not mandatory. However disposing of it forces two coincidences of scale to make $\mu_H^2$ sensitive to the SM Higgs vev. 

To show this we can write a left-right symmetric model which is approximately invariant under $H_1 \leftrightarrow H_2$ as in~\cite{Espinosa:2015eda, Dvali:2003br}. If $B\mu \lesssim 16 \pi^2 v^2$ and $\lambda_{6, 7} \lesssim 16 \pi^2 v^2/M^2_*$ , $\mu_H^2$ is dominated by the tree-level contributions from the vevs. Furthermore, the exchange symmetry forces $|m_{1,2}^2|\simeq v^2$ when $\mu_H^2 \simeq v^2$, just what we want to select the weak scale from $H_1 H_2$. Nonetheless, to make this model compatible with present LHC constraints we need both $||m_1^2|-|m_2^2||\gtrsim v^2$ and $B\mu \gtrsim v^2$. As we have just shown, to make $H_1 H_2$ a good trigger we have upper bounds of the same order on both quantities: 1) we do not want loop corrections to $\mu_H^2$ to dominate on the vevs, hence $B\mu \lesssim 16 \pi^2 v^2$, 2) we can not take the two masses too far apart, since breaking too much the exchange symmetry can lead to $\langle H_1 H_2 \rangle \simeq B\mu \frac{|m_1^2|}{|m_2^2|}$, which can be close to the weak scale even when $m_2^2 \simeq - m_1^2 \simeq M^2_*$. In summary we need both $||m_1^2|-|m_2^2||$ and $B\mu$ to be of $\mathcal{O}(v^2)$. So this is still an interesting possibility to consider, but it is not as simple as imposing the $\mathbb{Z}_2$ symmetry.

\section{The Standard Model Trigger}\label{sec:GGtrigger}

We now consider the SM trigger, expanding the discussion of \cite{TitoDAgnolo:2021nhd}. We take $\phi_\pm$ to have an axion-like coupling to gluons
\be
V_{G \phi} =-\frac{1}{32\pi^2}\left(\frac{\phi_+}{F_+}+ \frac{\phi_-}{F_-}+\theta\right){\rm Tr}[G\widetilde G]\, . \label{eq:gg}
\ee
For $m_{u,d} \lesssim 4 \pi f_\pi$, if we rotate $\phi_\pm$ in the quark mass matrix and match to the chiral Lagrangian at low energy, Eq.~\eqref{eq:gg} gives
\be
V_{G \phi} = - m_\pi^2 f_\pi^2 \sqrt{1-\frac{4 m_u m_d}{(m_u+m_d)^2}\sin^2\left(\frac{\phi_+}{2 F_+}+ \frac{\phi_-}{2 F_-}+\frac{\theta}{2}\right)} \simeq \frac{\Lambda^4(\langle h \rangle)}{2} \left(\frac{\phi_+}{F_+}+ \frac{\phi_-}{F_-}+\theta\right)^2  \!\!, \:\;\;\:\;\;\label{eq:phiH}
\ee
where the potential is switched on at the QCD phase transition by chiral symmetry breaking
\be
\Lambda^4(\langle h \rangle) = m_\pi^2 f_\pi^2 \frac{m_u m_d}{(m_u+m_d)^2} \,.
\ee
We stress that its size is a monotonic function of the Higgs vev even in the regime $m_{u,d} \gtrsim 4 \pi f_\pi$, although the functional form of $\Lambda(\langle h \rangle)$ becomes different. For the moment, we assume that the vacuum angle $\theta$ (which includes the quark-mass phases) is fixed and small because of some UV-mechanism that solves the strong-CP problem. Later, in Section \ref{sec:withStrongCP} we will relax this assumption and show that the mechanism can actually also solve the strong-CP problem by itself in a novel way~\cite{TitoDAgnolo:2021nhd}, if the $\theta$ angle instead scans in the landscape.

We consider a scalar potential with the same form as in section~\ref{sec:mechanism}:
\be
V_\pm(\phi_\pm)=m_{\pm}^2 M^2_\pm \left(\frac{\phi_\pm}{M_\pm} +\frac{\phi^2_\pm}{2 M^2_\pm} \pm \frac{\phi^3_\pm}{3 M^3_\pm} + \frac{\delta}{4} \frac{\phi^4_\pm}{M^4_\pm}\right)+...\label{eq:Vpm}
\ee
with the total potential being $V = V_+ + V_- + V_{G\phi}$. Notice that \eqref{eq:phiH} does not impose constraints on the naturalness of $V_\pm$. Therefore, $M_\pm$ are not related to the Higgs cutoff, which can be arbitrarily large. We take $M_\pm/F_\pm \ll 1$ so that in the local region of the potential $V_{G\phi}$ is dominated by the quadratic term in the second equality of Eq.~\eqref{eq:phiH}. As we will see in the following this is required by current measurements of the QCD $\theta$-angle.

We assume $m_+ \lesssim H(\Lambda_{\rm QCD})$, so that when $\phi_+$ starts to move,  from the local region $\phi_+ \sim M_+$ (otherwise the patch crunches anyway), the potential \eqref{eq:phiH} is already switched on.  Then, the $\phi_+$ potential is locally stabilized only if $\langle h \rangle > \mu_S$, with 
\be
\Lambda^4_S \equiv \Lambda^4(\langle \mu_S \rangle) \simeq m_+^2 F_+^2 \, . \label{eq:LambdaS}
\ee
This can be understood as follows: in absence of $V_{G\phi}$, $\phi_+$ does not have a metastable minimum in the local region $|\phi_+|\lesssim M_+$. In this region $V_{G\phi}$ is given approximately by the quadratic term in the second equality of Eq.~\eqref{eq:phiH}. The only monomial that can generate a minimum is the $\phi_+^2$ term in $V_{G\phi}$. The minimum is generated only if this term (with positive sign) dominates within $|\phi_+|\lesssim M_+$.

In general we may not have $\mu_S \simeq v$, so in this section we give formulas valid also for $\mu_S < v$, which is enough to select successfully the weak scale. For instance, in this case the size of local stable region around the metastable minimum is increased from $M_+$, at $\mu_S = v$, to \footnote{This formula is valid if the instability is generated by a cubic term, as in \eqref{eq:Vpm}. If, instead, the instability is generated by a quartic coupling, like in the example potential of \cite{TitoDAgnolo:2021nhd}, we find $\widetilde M_+ \simeq M_+ \Lambda^2_{\rm QCD}/\Lambda_S^2$.} 
\be
\widetilde{M}_+ \simeq \frac{\Lambda^4_{\rm QCD}}{\Lambda^4_S} M_+ \,.
\ee
The physical mass of the scalar is
\be
m_{\phi_+}^2 \simeq \frac{\Lambda^4_{\rm QCD}}{F_+^2} \simeq \frac{\Lambda^4_{\rm QCD}}{\Lambda^4_S} m_+^2 \,. \label{eq:mphiplus}
\ee

The above arguments show how we get a lower bound on the weak scale.
An upper bound  is generated as long as the $\phi_-$ potential in \eqref{eq:phiH} is dominated by the tadpole, i.e. $M_-/F_- \lesssim \widetilde M_+/F_+ + \theta$. In this case, the safe local minimum exists as long as $\langle h \rangle < \mu_B$, with
\be \label{eq:muB}
\Lambda_B^4 \equiv \Lambda^4(\langle \mu_B \rangle) \simeq \frac{m_-^2 M_- F_-}{\theta + \widetilde{M}_+ / F_+} . \label{eq:LambdaB}
\ee 
If both $\phi_+$ and $\phi_-$ exist in Nature, the only patches that do not crunch are those with $\mu_S < \langle h \rangle < \mu_B$. Given that, typically, large Higgs masses are favoured in the landscape, we have $v \approx \mu_B$, so that $\Lambda_B \approx \Lambda_{\rm QCD}$. The physical scalar mass is $m_{\phi_-} \simeq m_-$. This could be smaller or bigger than $m_+$ and $H(\Lambda_{\rm QCD})$. Accordingly, during $\phi_-$ dynamics the other scalar $\phi_+$ could be still frozen by Hubble friction or not. 
We have replaced the unknown $\phi_+$ misalignment at the time when $\phi_-$ starts to move, with its typical value $\widetilde M_+$, i.e. the size of the local stability region close to the safe metastable minimum of $\phi_+$. Notice that $\phi_+$ moves by an amount $\sim \widetilde{M}_+$ after the QCD phase transition, so even patches for which the denominator in \eqref{eq:muB} is initially tuned to be small can  survive until today only if $\langle h \rangle \lesssim \mu_B$, since the denominator will effectively be detuned when $\phi_+$ starts to move.  The $\theta$-angle today is 
\be
\theta_0 \simeq \theta + \frac{\widetilde{M}_+}{F_+} + \frac{M_-}{F_-} \simeq \theta + \frac{\widetilde{M}_+}{F_+}  \,.
\ee
We had already assumed $\theta \lesssim 10^{-10}$ from an unspecified UV solution to the strong CP problem (for instance of the Nelson-Barr type~\cite{Nelson:1983zb,Barr:1984qx}), we further require $\widetilde{M}_+/F_+ \lesssim \theta_{\rm exp} \simeq 10^{-10}$. 

Notice that along the flat direction of \eqref{eq:phiH}, $F_- \phi_+= -F_+ \phi_- - \theta F_+ F_-$, the potential is not sensitive to the Higgs vev. However, with our assumptions (and at fixed $\theta$), generically the flat direction does not intersect the local stability region $\phi_\pm \sim M_\pm$ and hence it does not pose a threat to the mechanism. 

\subsection{Solving also the strong-CP problem}\label{sec:withStrongCP}

So far we have assumed that the $\theta$ angle is set to be small by some unspecified mechanism operating at a high energy scale ($E\gg \sqrt{m_{\phi_\pm} M_\pm}$). We now show that the usual Peccei-Quinn solution is not compatible with the mechanism. Let us assume that an axion $a$ is present, heavier than $\phi_\pm$, so that \eqref{eq:phiH} is modified to
\be
V_{G \phi a} \simeq  \frac{\Lambda^4(\langle h \rangle)}{2} \left(\frac{\phi_+}{F_+}+ \frac{\phi_-}{F_-}+ \frac{a}{f}\right)^2 \;, 
\ee
having used the shift-symmetry of $a$ to absorb the UV $\theta$ angle. Then, the first scalar that starts rolling is the axion itself, which rapidly relaxes the whole $\langle h \rangle$-dependent potential to 0; this is continuously readjusted to 0 even subsequently, during the slower motion of $\phi_\pm$. As a consequence, $\phi_\pm$ would not be sensitive to the Higgs vev. Notice that some small Peccei-Quinn breaking  potential for the axion, coming from the UV, would not help, being independent on the Higgs vev.

However, if the $\theta$-angle is also scanned in the landscape (for instance because of the presence of a scalar coupled to $G \widetilde G$ and lighter than $\phi_\pm$), then \textit{our mechanism itself solves the strong-CP problem in a novel way}, in addition to the Higgs hierarchy problem~\cite{TitoDAgnolo:2021nhd}. This occurs because the $\phi_+$ metastable minimum is generated only if $\theta$ is small enough that the minimum of $\eqref{eq:phiH}$ lies within the local region $\phi \sim M_+$, where the destabilizing cubic term of $V_+$ does not dominate. Otherwise the patch crunches, in the same way as the ones with a ``wrong'' value of the Higgs vev. A small $\theta$ is selected by this requirement:
\be
\frac{\widetilde M_+}{F_+} \gtrsim \theta \;.
\ee
The only patches that do not crunch are those with $\mu_S \lesssim \langle h \rangle \lesssim \mu_B$ \textit{and} $\theta_0 \ll 1$. This novel solution to the strong-CP problem has its own phenomenological features that distinguish it clearly from the axion one, as discussed in~\cite{TitoDAgnolo:2021nhd} and summarized in the next subsection. Additionally, the same dynamics selects a small and nonzero Higgs vev.

Before discussing the phenomenology, we point out a subtlety that arises once $\theta$ scans in the landscape. In this case, there certainly exist patches with tuned values of $\theta$ such that the flat direction of \eqref{eq:phiH} crosses the local stability region $\phi_\pm \sim M_\pm$ and therefore becomes relevant. Recall that along the flat direction the potential is not sensitive to the Higgs vev. 
On the one hand, it is possible to show that the potential along this direction is locally stabilized by the quadratic terms of \eqref{eq:Vpm}, as long as $\Lambda_B \gtrsim \Lambda_S$. 
On the other hand, this ``tuned'' local minimum keeps being present in the potential even for large values of $\Lambda(\langle h \rangle) \gg \Lambda_{\rm QCD}$, threatening the successful selection of $\langle h \rangle \lesssim v$: as just mentioned along the flat direction the potential is locally stable to start with, along the orthogonal direction it is made stable by the large contribution of \eqref{eq:phiH}. However, our mechanism is still successful because these metastable patches with $\Lambda(\langle h \rangle) \gg \Lambda_{\rm QCD}$ are \textit{doubly tuned}. First, in order for the flat direction to cross the local stability region, $\theta$ needs to be tuned by an amount 
\be
\epsilon_\theta \sim \frac{M_-/F_-}{\widetilde{M}_+/F_+} \ll 1
\ee
as compared to stable patches with $\Lambda(\langle h \rangle) \sim \Lambda_{\rm QCD}$. Second, given that the flat direction is essentially parallel to $\phi_- \sim \rm const.$, the barrier along it is $\Delta V_\parallel \sim m_-^2 M_-^2$. Then, for these ``bad'' patches to be metastable, the initial value of $\phi_+$ needs to be tuned to lie within a tiny region of size $\Delta \phi_+$ such that $\Delta V_\perp \lesssim \Delta V_\parallel$, otherwise the combined evolution of $\phi_\pm$, which explores the phase-space energetically allowed, would probe the instability. This gives an additional tuning $\epsilon_{\phi_+} \sim {\Delta \phi_+}/{\widetilde M_+}$, with:
\be
\frac{\Lambda(\langle h \rangle)^4 \Delta \phi_+^2}{F_+^2} \sim m_-^2 M_-^2 \,,
\ee
yielding
\be
\epsilon_{\phi_+} \sim \sqrt{\frac{M_-/F_-}{\widetilde{M}_+/F_+} } \frac{\Lambda_B^2}{\Lambda(\langle h \rangle)^2 } \ll 1\,.
\ee
The combined tuning $\epsilon_\theta \epsilon_{\phi_+}$ can be made arbitrarily small by taking $M_-/F_- \ll \widetilde{M}_+/F_+$, so to compensate any reasonable a priori preference for large values of $\Lambda(\langle h \rangle)$ in the landscape, thus making the doubly tuned patches irrelevant, being arbitrarily rare or absent altogether\footnote{This latter possibility happens in case the tuned initial values of $\theta$ or $\phi_+$ are forbidden by additional interactions in the UV, for instance non-minimal couplings to gravity during inflation. }.

\subsection{Smoking-gun phenomenological pattern}
The cosmology of the model is basically the same as for the $H_1 H_2$ trigger, with the role of the electroweak phase transition replaced by the QCD one. In particular, the scalar $\phi_+$ needs to be lighter than $H(\Lambda_{\rm QCD})$, or the universe would crunch independently of $\langle h \rangle$ before the Higgs-dependent potential is switched on. 

\begin{figure}
$$\includegraphics[width=0.4\textwidth]{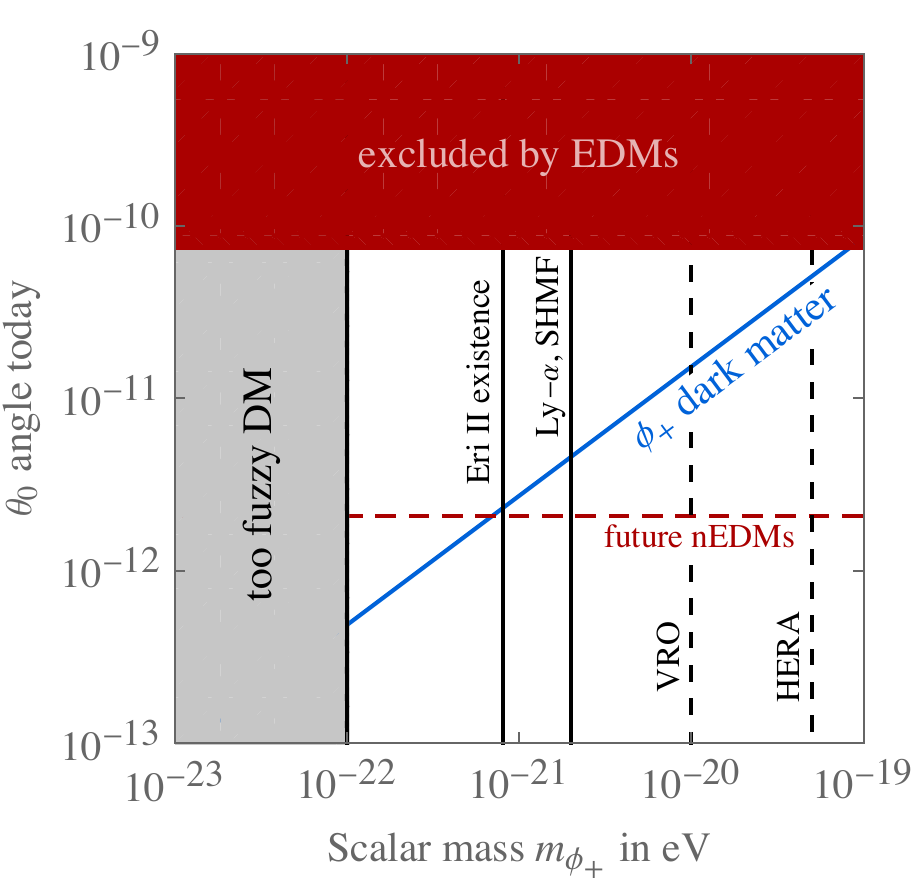} \qquad\qquad \includegraphics[width=0.4\textwidth]{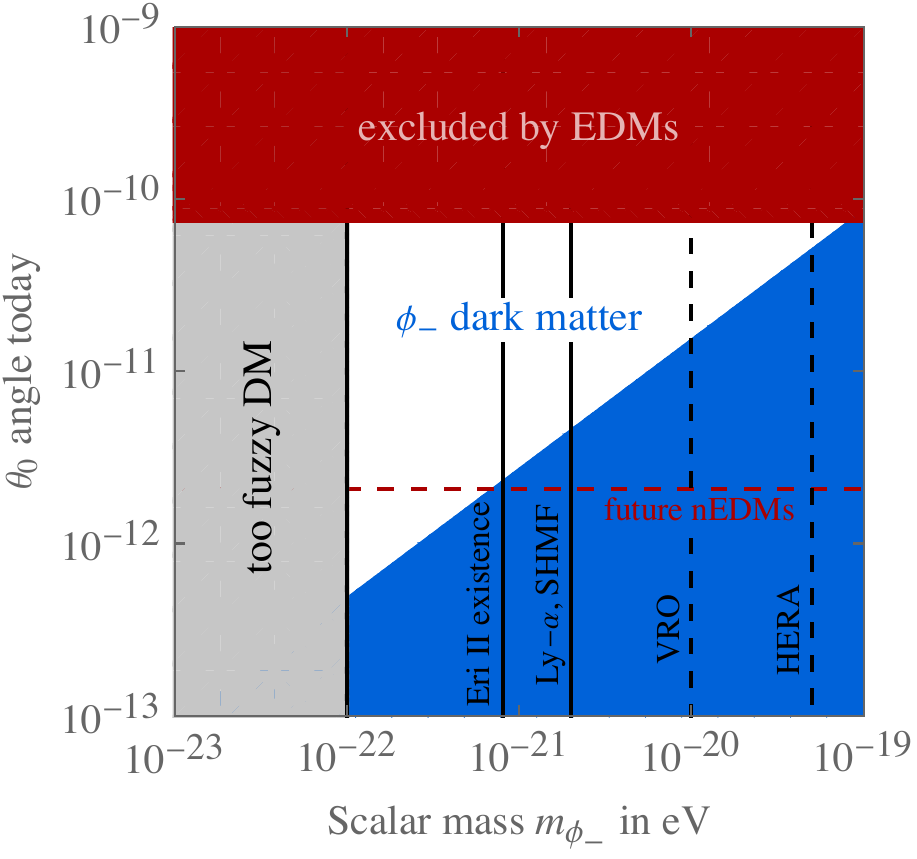}$$
\caption{Parameter space for  which $\phi_+$ (left panel) or $\phi_-$ (right panel) constitute the totality of DM of the Universe, as function of their mass and the $\theta$-angle today. The DM relic density is reproduced along the blue line (left panel) or white region (right panel). The red shaded region (dashed line) shows bounds~\cite{Abel:2020gbr} (future prospects~\cite{Abel:2018yeo,SNSnEDM,Filippone:2018vxf}) from hadronic EDM searches. New ideas involving molecular compounds could further improve future sensitivities~\cite{Hutzler:2020lmj, PhysRevA.100.032514, doi:10.1063/1.5098540, PhysRevC.94.025501, Yu:2020wtz}.
 We also plot in black constraints on fuzzy DM from 
Lyman-$\alpha$ forest~\cite{Leong:2018opi, Irsic:2017yje, Kobayashi:2017jcf, Armengaud:2017nkf, Bozek:2014uqa, Zhang:2017chj}, measurements of the subhalo mass function~\cite{Schutz:2020jox} and the Eridanus II dwarf galaxy~\cite{Marsh:2018zyw} (similar to the constraints from other dwarf galaxies~\cite{Safarzadeh:2019sre, Bar:2018acw}). We shade the area where multiple observations disfavor the corresponding DM mass hypotheses~\cite{Hui:2016ltb}. The dashed lines denotes the potential sensitivity from future observations in 21 cm cosmology (HERA)~\cite{Munoz:2019hjh} and by the Vera Rubin Observatory~\cite{Drlica-Wagner:2019xan}.  \label{fig:SMtriggerDM}}
\end{figure}

Both scalars can constitute the totality of dark matter in the Universe, yielding a DM phenomenology cross-correlated with EDM experiments, as studied in detail in \cite{TitoDAgnolo:2021nhd} for $\phi_+$. Let us start from this scalar. At the QCD phase transition it gets a kick of order $\widetilde{M}_+$, which dominates its oscillations. Then, its energy density when it starts oscillating after the QCD transition is $\rho_+ \sim m_{\phi_+}^2 \widetilde{M}_+^2 \sim \theta_0^2 \Lambda^4_{\rm QCD}$, giving the relic density today
\be
\frac{\rho_{\phi_+}}{\rho_{\rm DM}} 
\simeq \frac{\theta_0^2 \Lambda^4_{\rm QCD}}{T_{\rm eq} M_{\rm Pl}^{3/2} m_{\phi_+}^{3/2}}\simeq \left(\frac{\theta_0}{10^{-10}}\right)^2\left(\frac{10^{-19}\;{\rm eV}}{m_{\phi_+}}\right)^{3/2} \, .\label{eq:RDplus}
\ee
Therefore, its relic density is $\simeq \theta_0^2$ times smaller than the one of a Peccei-Quinn axion with the same mass, avoiding overclosure constraints on light axions. Also $\phi_-$ can be the dark matter of the Universe, if light enough. Analogously to $\phi_+$, its energy density at the onset of its oscillations is $\rho_- \sim m_{-}^2 {M}_-^2 \sim (M_-/F_-) \theta_0 \Lambda^4_{\rm QCD} \lesssim \theta_0^2 \Lambda^4_{\rm QCD}$, smaller than the one for $\phi_+$. However, it can give the correct relic density  if lighter than $\phi_+$:
\be
\frac{\rho_{\phi_-}}{\rho_{\rm DM}} 
\simeq \frac{\theta_0  \Lambda^4_{\rm QCD} M_-/F_-}{T_{\rm eq} M_{\rm Pl}^{3/2} m_{\phi_-}^{3/2}}\simeq \left(\frac{\theta_0}{10^{-10}}\right) \left(\frac{M_-/F_-}{10^{-10}}\right) \left(\frac{10^{-19}\;{\rm eV}}{m_{\phi_-}}\right)^{3/2} \, .\label{eq:RDminus}
\ee

Summarizing, the scalar $\phi_+$ is an axion of mass $m_{\phi_+} \lesssim 10^{-11} \, \text{eV}$ which lies on the QCD line $m_{\phi_+} \simeq \Lambda_{\rm QCD}^2/F_+$, as it can be seen by combining \eqref{eq:LambdaS} and \eqref{eq:mphiplus}. Instead, $\phi_-$ is an ALP with a mass comparable to or larger than a QCD axion with the same couplings, as it can be seen from \eqref{eq:LambdaB} and $M_-/F_- \lesssim \theta_0$.

Notice that $\phi_\pm$ do  not give rise to black hole superradiance in the region $m_{\phi_\pm} \sim 10^{-12} \, \text{eV}$ because of the self-coupling in Eq.~\eqref{eq:Vpm}~\cite{Baryakhtar:2020gao}. If either of them is observed in this region, this would then constitute a first characteristic trait that distinguishes our scalars from the Peccei-Quinn axion. 

However, the best phenomenological prospects occur if they are lighter and constitute the dark matter of the Universe, as shown in Figure \ref{fig:SMtriggerDM}. Their relic density is strongly correlated with the value of the $\theta$ angle today. This is a 1-to-1 correspondence for $\phi_+$, while for $\phi_-$ there is an additional parameter $M_-/F_-$. However this ratio has the upper bound $M_-/F_-\lesssim \theta_0$.
As a consequence, limits on  fuzzy DM imply $\theta_0 \gtrsim 10^{-12}$, observable at future EDM experiments~\cite{Abel:2018yeo,SNSnEDM,Filippone:2018vxf}: if either $\phi_+$ or $\phi_-$ is dark matter, we predict sizeable EDMs. A joint observation of $\theta_0$ in the near future and a measurement of the DM mass would allow to test the smoking-gun relations in Eq.~\eqref{eq:RDplus} or~\eqref{eq:RDminus}. A combination of future EDM measurements and fuzzy DM probes~\cite{Safarzadeh:2019sre, Bar:2018acw, Leong:2018opi, Irsic:2017yje, Kobayashi:2017jcf, Armengaud:2017nkf, Bozek:2014uqa, Zhang:2017chj, Hui:2016ltb,  Schutz:2020jox, Marsh:2018zyw,Munoz:2019hjh, Drlica-Wagner:2019xan} can fully test the hypothesis of $\phi_\pm$ DM, as shown in Fig.~\ref{fig:SMtriggerDM}.

\section{Conclusions}\label{sec:conclusion}
The two main discoveries of the LHC so far have been: the Higgs boson and the unnaturalness of its mass. In this work we have presented a novel mechanism that explains this unnaturalness by means of cosmological selection: the multiverse is populated by patches with different values of the Higgs mass; the ones where the EW scale is too small or too large crunch in a short time, the other ones, with the observed (unnaturally small) value of the EW scale survive and expand cosmologically, resulting in an universe as the one that we observe. In a companion paper~\cite{TitoDAgnolo:2021nhd} we called this scenario \textit{Sliding Naturalness}, since the crunching is due to two light scalars sliding down their potential.

The phenomenology of our proposals depends strongly on the trigger operator that connects the two scalars to the SM. For the $H_1 H_2$ trigger, as discussed in detail in~\cite{Arkani-Hamed:2020yna} and summarized in Section~\ref{sec:H1H2trigger}, the most favourable prospects for detection come from the observation of the type-0 2HDM at colliders, with the high-luminosity LHC probing completely this possibility.  For the SM trigger $G \widetilde G$, the mechanism yields ALP phenomenology. However, a remarkable feature of our scenario~\cite{TitoDAgnolo:2021nhd} is that in this case it also solves automatically the strong-CP problem, in a novel way, different from the usual Peccei-Quinn mechanism, as described in Section~\ref{sec:GGtrigger}. 

In both cases, the oscillations of the two scalars can constitute the totality of dark matter in the Universe (see Section~\ref{sec:pheno}). In the case of the SM trigger, this possibility additionally implies a large value of the QCD  angle $\theta \gtrsim 10^{-12}$, observable in the near future, and strongly correlated to the DM mass, the latter in the fuzzy-DM range (see Figure~\ref{fig:SMtriggerDM}).

In the last years several cosmological approaches to naturalness have been developed. In the preliminary discussion of Section~\ref{sec:CN} we have attempted to draw an unified picture by identifying the general features of these proposals and then focused on what we called dynamical selection. This class of models is the one with the best prospect of detection. We identified three main ingredients. First, the presence of a landscape for the Higgs mass, which is often difficult to observe. Second, the presence of light scalars coupled to the SM. By means of NDA considerations, we argued that their lightness is related to having a large cutoff for the Higgs sector. While the presence of light scalars is not common to all cosmological approaches to the hierarchy problem, it is frequent enough to provide guidance for experimental searches. Third, a trigger operator~\cite{Arkani-Hamed:2020yna} that connects the scalars to the SM, which determines the phenomenological strategy to probe these solutions.

There are a number of important features that single out our mechanism as compared to other existing proposals in the literature.
First, an important distinction between models of cosmological naturalness arises from how they influence inflation. In some cases the Hubble rate during inflation is required to be smaller than $m_h$ and an exponentially large number of $e$-folds might be needed. This clearly requires additional model building that the reader is screened from, but which might considerably complicate the model or introduce tuning. Our mechanism, instead, factorizes from the sector responsible for inflation. Second, the model can have large cutoffs (comparable to $M_{\rm Pl}$) for both the CC and Higgs mass and at low energy only predicts two extremely weakly coupled scalars with a simple potential. Finally, as argued in~\cite{TitoDAgnolo:2021nhd}, Sliding Naturalness is compatible with modern swampland conjectures and does not suffer from ambiguities connected to eternal inflation.\footnote{More precisely, our mechanism is compatible with eternal inflation (but does not require it, as long as the landscape is populated by some mechanism), and at the same time it does not suffer from the so-called measure problem, since the relevant dynamics that selects the Higgs mass takes place after reheating, at the EW or QCD phase transition.}

We cannot know if the unnaturalness of the Higgs mass discovered by the LHC will ultimately be explained by cosmological dynamics. However, the progress of the last years gives us a plausible alternative to traditional solutions to the problem or to accepting tuning. This framework can be tested experimentally in the next decade. In this context, the novel mechanism that we propose is, in our opinion, a particularly attractive solution, in view of its simplicity, and compatibility with simple realizations of other sectors of the theory. 

\section*{Acknowledgments}

We thank R. Rattazzi, N. Arkani-Hamed, and H.D. Kim for useful discussions.


\bibliography{refs}
\bibliographystyle{utphys}


\end{document}